\documentclass[aps,prd,amsmath,amssymb,10pt,letterpaper,balancelastpage,showpacs,superscriptaddress,nofootinbib,floatfix,notitlepage,longbibliography,twocolumn
]{revtex4-2}
\usepackage{comment}
\usepackage{graphics}
\usepackage{bm}
\usepackage{graphicx}
\usepackage{color}
\usepackage[dvipsnames]{xcolor}
\usepackage{amsmath}
\usepackage{amssymb}
\usepackage{mathrsfs}
\usepackage{graphicx}
\usepackage{ragged2e}
\usepackage{caption}
\usepackage{subcaption}
\DeclareCaptionJustification{justified}{\justifying}
\usepackage{accents}
\usepackage{adjustbox}
\usepackage{multirow}
\usepackage{placeins} 
\usepackage{stackengine} 
\usepackage{float} 
\usepackage{ulem}
\usepackage[
    starfontserif 
    ]{starfont}
\usepackage[breaklinks=true]{hyperref}
\usepackage{xurl} 
\hypersetup{
    colorlinks,
    citecolor=black,
    filecolor=black,
    linkcolor=black,
    urlcolor=black
}

\DeclareSymbolFont{starfontsym}{OT1}{sts}{m}{n}
\DeclareMathSymbol{\mathSun}{\mathord}{starfontsym}{115}
\DeclareMathSymbol{\mathMercury}{\mathord}{starfontsym}{102}
\DeclareMathSymbol{\mathVenus}{\mathord}{starfontsym}{103}
\DeclareMathSymbol{\mathTerra}{\mathord}{starfontsym}{76}
\DeclareMathSymbol{\mathvarTerra}{\mathord}{starfontsym}{108}
\DeclareMathSymbol{\mathMoon}{\mathord}{starfontsym}{100}
\DeclareMathSymbol{\mathvarMoon}{\mathord}{starfontsym}{97}
\DeclareMathSymbol{\mathMars}{\mathord}{starfontsym}{104}
\DeclareMathSymbol{\mathJupiter}{\mathord}{starfontsym}{106}
\DeclareMathSymbol{\mathSaturn}{\mathord}{starfontsym}{83}
\DeclareMathSymbol{\mathUranus}{\mathord}{starfontsym}{70}
\DeclareMathSymbol{\mathvarUranus}{\mathord}{starfontsym}{65}
\DeclareMathSymbol{\mathNeptune}{\mathord}{starfontsym}{71}
\DeclareMathSymbol{\mathPluto}{\mathord}{starfontsym}{74}
\DeclareMathSymbol{\mathvarPluto}{\mathord}{starfontsym}{72}

\newcommand{\numcolor}[1]{\textcolor{black}{#1}}

\newcommand{\eq}[1]{eq.~(#1)}
\newcommand{\Eq}[1]{Eq.~(#1)}

\newcommand{\rbold}{\mathbf{r}}

\newcommand{\vbold}{\mathbf{v}}

\newcommand{\Gg}{\Gamma_{\gamma}}
\newcommand{\sxN}{\sigma_{\Ht n}}
\newcommand{\sxA}{\sigma_{\Ht A}}
\newcommand{\GxA}{\Gamma_{\Ht A}}
\newcommand{\Veff}{V_{\text{eff}}}
\newcommand{\nx}{n_{\Ht}}
\newcommand{\vmin}{v_{\text{min}}}
\newcommand{\vglobalmin}{v_{\text{min,global}}}
\newcommand{\mx}{m_{\Ht}}
\newcommand{\vmax}{v_{\text{max}}}
\newcommand{\vmaxSHM}{v^{\text{SHM}}_{\text{max}}}
\newcommand{\vmaxLMC}{v^{\text{LMC}}_{\text{max}}}
\newcommand{\vesc}{v_{\text{esc}}}

\newcommand{\Gammasignal}{\Gamma_{\text{signal}}}

\newcommand{\Vpb}{V_\text{Pb}}

\newcommand{\npb}{n_\text{Pb,supp}}

\newcommand{\Rpb}{R_\text{Pb}}
\newcommand{\Veffpb}{V_{\text{eff,Pb}}}
\newcommand{\deltamaxA}{\delta_{\text{max},A}}
\newcommand{\deltamaxALMC}{\delta^{\text{LMC}}_{\text{max},A}}
\newcommand{\deltamaxLMC}[1]{\delta^{\text{LMC}}_{\text{max},\text{#1}}}
\newcommand{\deltamaxASHM}{\delta^{\text{SHM}}_{\text{max},A}}
\newcommand{\muA}{\mu_A}
\newcommand{\kms}{\text{km}/\text{s}}
\newcommand{\Msun}{\mathrm{M}_{\mathSun}}
\newcommand{\REarth}{\mathrm{R}_{\mathTerra}}
\newcommand{\fdet}{f^{\text{det}}}
\newcommand{\lmaxA}{l_{\text{max},A}}
\newcommand{\lmaxALMC}{l^{\text{LMC}}_{\text{max},A}}
\newcommand{\lmaxLMC}[1]{l^{\text{LMC}}_{\text{max,#1}}}
\newcommand{\lmaxASHM}{l^{\text{SHM}}_{\text{max},A}}
\newcommand{\hmax}{h_{\text{max}}}

\newcommand{\deltaminA}{\delta_{\text{min},A}}
\newcommand{\deltaminFe}{\delta_{\text{min,Fe}}}
\newcommand{\Ethr}{E_{\text{thr}}}
\newcommand{\ttotal}{t_{\text{total}}}
\newcommand{\tsignal}{t_{\text{signal}}}
\newcommand{\Gammabackground}{\Gamma_{\text{bg}}}

\newcommand{\Ht}{\tilde{H}}
\newcommand{\Hone}{\tilde{H}_1}
\newcommand{\Htwo}{\tilde{H}_2}
\newcommand{\lH}{l_{\tilde{H}_2}}

\begin{document}
\title{Enhancing direct detection of Higgsino dark matter}
\author{Peter W.~Graham}
\email{pwgraham@stanford.edu}
\affiliation{Stanford Institute for Theoretical Physics, Department of Physics, Stanford University, Stanford, California
94305, USA}
\affiliation{Kavli Institute for Particle Astrophysics and Cosmology, Department of Physics, Stanford University, Stanford, California 94305, USA}
\author{Harikrishnan Ramani}
\email{hramani@udel.edu}
\affiliation{Stanford Institute for Theoretical Physics, Department of Physics, Stanford University, Stanford, California 94305, USA}
\affiliation{Department of Physics and Astronomy, University of Delaware, Newark, Delaware 19716, USA}
\author{Samuel S. Y. Wong}
\email{samswong@stanford.edu}
\affiliation{Stanford Institute for Theoretical Physics, Department of Physics, Stanford University, Stanford, California 94305, USA}

\begin{abstract} 

While much supersymmetric weakly interacting massive particle (WIMP) parameter space has been ruled out, one remaining important candidate is Higgsino dark matter.
The Higgsino can naturally realize the ``inelastic dark matter" scenario, where the scattering off a nucleus occurs between two nearly-degenerate states, making it invisible to WIMP direct detection experiments if the splitting is too large to be excited.
It was realized that a ``luminous dark matter" detection process, where the Higgsino upscatters in the Earth and subsequently decays into a photon in a large neutrino detector, offers the best sensitivity to such a scenario.
We consider the possibility of adding a large volume of a heavy element, such as Pb or U, around the detector.
We also consider the presence of U and Th in the Earth itself, and the effect of an enhanced high-velocity tail of the dark matter distribution due to the presence of the Large Magellanic Cloud.
These effects can significantly improve the sensitivity of detectors such as JUNO, SNO+, KamLAND, and Borexino, potentially making it possible in the future to cover much of the remaining parameter space for this classic supersymmetric WIMP dark matter.
\end{abstract}

\maketitle


\section{Introduction}

The nature of dark matter (DM) is one of the greatest outstanding questions in particle physics and cosmology.  A weakly interacting massive particle (WIMP) is one of the best-motivated DM candidates, and supersymmetry (SUSY) provides classic WIMP candidates.  However, WIMP dark matter and SUSY have been searched for extensively in collider, direct, and indirect detection experiments and much of WIMP and SUSY parameter spaces have been ruled out.

One interesting remaining possibility for a WIMP is the inelastic dark matter scenario \cite{Hall:1997ah, Tucker-Smith:2001myb}.  In this scenario, there is a small mass splitting $\delta$ between two nearly degenerate states, and scattering off a nucleus transitions one state to the other.  The signal in a direct detection experiment may then be suppressed if there is insufficient energy in the DM-nucleus collision to excite this splitting.  While traditional DM direct detection experiments constrain the case of small $\delta$, a sufficiently large $\delta$ will evade these constraints.  As a result, there is still a significant amount of interesting, allowed parameter space for WIMP dark matter if it has this small splitting.  Furthermore, SUSY can naturally realize this inelastic dark matter scenario if the DM candidate is a Higgsino.  Excitingly, as we will discuss in Section~\ref{Sec: Higgsino DM}, much of this Higgsino DM parameter space remains unexplored.

The natural question then becomes how to detect such Higgsino DM with a scattering that must excite from the lighter to the heavier state.  Interestingly, the heavier Higgsino state decays to the lighter state by emitting a photon, with a decay length that falls within experimentally accessible length scales. 
 This then raises the following interesting possibility known as ``luminous dark matter" \cite{LuminousDM}.  The incoming DM particle can upscatter to the heavier state anywhere within the Earth and may then pass through a detector.  If it decays while inside the detector, the emitted photon can be detected.  Importantly, in the relevant parts of parameter space, the rate for this ``luminous" detection process is roughly the same as the rate for direct collisions in the detector volume, assuming the detector material is the same as the material in the Earth surrounding the detector \cite{LuminousDM}.  This case was considered in Ref.~\cite{Luminous}, where it was pointed out that large underground neutrino detectors, such as Borexino and JUNO, offer the best sensitivity to Higgsino DM.  A new search strategy for these detectors using daily modulation was proposed and the resulting sensitivity was calculated.
 
 Such a search \cite{Luminous} can cover up to higher $\delta$ because the Higgsino can scatter off of heavy elements in the Earth, such as lead (Pb), rather than the lighter elements, such as xenon (Xe), in traditional WIMP detectors.  The highest splitting $\delta$ that can be excited by a collision with a nucleus is set by the DM velocity and the reduced mass between the DM and the nucleus.  Since Higgsino DM is approximately $ 1.1 \, \text{TeV}$ in mass, the reduced mass is set by the nucleus mass.  Therefore, the reach in $\delta$ is improved by scattering off of heavier elements.  Although WIMP detectors tend not to have such heavy elements in them, using this luminous detection process allows us to take advantage of the heavier elements in the Earth.

In this paper, we build on this previous work \cite{Luminous} to extend the reach for Higgsino DM to higher mass splittings $\delta$, thereby allowing access to more of the well-motivated supersymmetric parameter space.  In particular, we consider the effect of adding a large amount of heavy material, such as Pb or uranium (U), around a large neutrino detector such as JUNO, SNO+, or KamLAND.  This is similar to the proposal in Ref.~\cite{Pospelov:2013nea}, where DM scattering off lead shielding around a neutrinoless double beta decay detector was considered.  The addition of heavy material increases the amount of scattering since this introduces a high density of such heavy elements close to the detector, instead of the lower average densities of these elements found naturally in the Earth.  We also calculate the sensitivity increase from including the naturally occurring amounts of U and thorium (Th) in the Earth in the usual luminous detection calculation.  Although these elements are present in low densities in the Earth, they are the heaviest available elements and thus help boost the sensitivity at the highest values of $\delta$. 
 Additionally, we consider the possibility that the DM population in the Milky Way (MW) has an enhanced high-velocity tail due to the presence of the Large Magellanic Cloud (LMC).  Although this represents a small fraction of the total DM, such a high-velocity population has an outsized impact on the detection of Higgsino DM because the reach in $\delta$ is limited by the energy available, which depends quadratically on the incoming DM velocity.  We also recalculate the existing limits on inelastic DM from traditional WIMP experiments using this new velocity distribution.
We find that both adding heavy material around the detectors and considering the effect of a possible high-velocity tail can significantly enhance the sensitivity of large neutrino detectors to Higgsino DM.

\section{Higgsino Dark Matter}
\label{Sec: Higgsino DM}

\begin{figure*}[t]
    \centering
    
        \centering
        \includegraphics[width=0.9\textwidth]{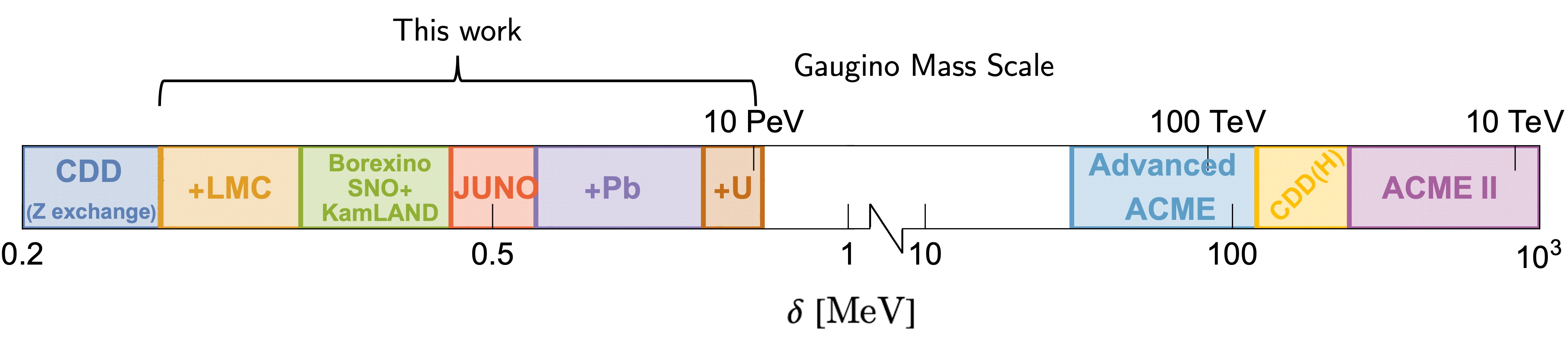}
    \caption{Summary of existing limits and projections on the mass splitting $\delta$ for the Higgsino dark matter scenario for $\mx=1.1$~TeV. Existing limits come from tree-level direct detection via Z-exchange~\cite{PandaX4T2021,PICO2023} in dark blue for small splitting and ACME II~\cite{acme2018improved} in purple. The bino and wino masses are assumed to be equal.}
    \label{fig:graphic}
    
\end{figure*}

The Higgsino arises as a natural DM candidate in the minimal supersymmetric Standard Model (MSSM) or more generic models of low-scale supersymmetry.  The Higgsino as a DM candidate has been well studied~\cite{LastWIMP}.  We provide a succinct summary of the important parameters here. 

The Higgsino has Dirac mass $\mu \tilde{H}_u \tilde{H}_d$ before electroweak symmetry breaking (EWSB), which leads to a neutral Dirac fermion and a charged Dirac fermion with identical masses. 
 However, EWSB results in a Majorana mass $m_M\approx m_Z^2\left(\frac{\sin^2\theta_W}{M_1}+\frac{\cos^2\theta_W}{M_2}\right)$, where $m_Z$, $M_1$, and $M_2$ are the masses of the Z boson, the bino, and the wino, respectively.  In the limit where the gauginos are much heavier than the Z boson, the Majorana mass is small compared to the Dirac mass $\mu$, resulting in pseudo-Dirac Higgsinos---i.e., two neutral Majorana Higgsinos $\tilde{H}_1$ and $\tilde{H}_2$ that are almost degenerate.  Their individual masses are $M_{\tilde{H}_{1,2}} \approx\mu \mp \frac{\delta}{2}$ with $\delta=m_M$, the Majorana mass generated after EWSB.  Therefore, the mass splitting is given by
\begin{align}
\delta \approx m_Z^2\left(\frac{\sin^2\theta_W}{M_1}+\frac{\cos^2\theta_W}{M_2}\right) ~.
\label{eq:splitting}
\end{align}

Particularly compelling is the Higgsino mass $\mu\approx 1.1 ~\textrm{TeV}$, as this would produce the correct relic abundance observed today via freeze-out.  The lifetime of the heavier Higgsino $\tilde{H}_2$ is typically much shorter than the age of the universe, so the lighter state $\tilde{H}_1$ constitutes all of dark matter today.  Whenever there is no ambiguity, we will denote the lighter Higgsino state by $\Ht$ and its mass by $\mx$.

The $1.1$~TeV Higgsino mass is well above the existing limits set by colliders~\cite{aaboud2018search}. There are promising prospects for its indirect detection~\cite{Rinchiuso:2020skh,Rodd:2024qsi} at CTA~\cite{CTAConsortium:2010umy}. However this critically depends on the amount of dark matter at the galactic center which is uncertain. We explore a complementary direction in this work: its direct detection on Earth. 

 The Higgsino splitting after EWSB results in a purely off-diagonal coupling of the Z boson to the Higgsino mass eigenstates. 
 While the splitting is negligible at freeze-out scales, it becomes crucial for direct detection today.  The Z mediator produces a per-nucleon cross section for direct detection
\begin{align} \label{eq:Higgsino cross section}
    \sigma_{\Ht n} \approx 10^{-39} \textrm{cm}^2 ~.
\end{align}
Despite being a large cross section for direct detection purposes, this process is inelastic in nature, as the lighter $\tilde{H}_1$ needs to up-convert to $\tilde{H}_2$.  This can only occur if there is enough energy in the center-of-momentum frame.  The largest splitting we can probe in collisional direct detection (CDD), $\delta_{\rm CDD}$, is given by  
\begin{align}
    \delta_{\rm CDD} =\frac{1}{2}\muA v_{\rm max}^2~,
    \label{eq:maxsplit}
\end{align} 
where $\muA \approx m_A$ is the reduced mass of the target nucleus with mass $m_A$ (and nuclear mass number $A$) that is scattering with the Higgsino $\tilde{H}$, and $v_{\rm max}$ is the maximum DM velocity locally.  Since $\frac{1}{2}\muA \vmax^2\approx 200~\textrm{keV}$ for a typical collisional direct detection nucleus and the Milky Way DM population, this sets the maximum splitting above which there are no current limits on the Higgsino.  This falls just short of the $\delta \gtrsim 200$~keV ($\delta \gtrsim 1~$ MeV) for small (large) $\tan \beta$ motivated by fine-tuning arguments~\cite{LastWIMP}. Recently, metastable isomers~\cite{isomer1,isomer2,isomer3,isomer4} have been used to set constraints on larger $\delta$ for generic inelastic splittings, but this strategy is not expected to reach the small cross sections relevant for the Higgsino.

Figure~\ref{fig:graphic} provides a rough summary of existing and future coverage for the splitting $\delta$ for $\mx$=$1.1$~TeV. We also show for reference the approximate gaugino mass derived from \Eq{\ref{eq:splitting}} on the top axis. Existing limits from CDD found in literature that go up to roughly $260$~keV~\cite{PandaX4T2021,PICO2023} are shown in blue and marked ``CDD (Z exchange)." 

Direct detection could also occur via elastic processes such as Higgs exchange or via one-loop. The Higgsino-Higgs coupling arises via its mixing with gauginos. As a result, the Higgs exchange cross section is dominant when the gaugino mass is close to the Higgsino mass and decouples with the gaugino mass. Current limits on the Higgs exchange rule out Higgsinos only if the gauginos are also at the TeV scale~\cite{LastWIMP}, which in turn constrains mass splittings of $\delta\approx 1~\textrm{GeV}$ or larger. The projected reach for the LZ experiment corresponds to 10-TeV gauginos, which in turn translates to splittings of $\delta\approx 100~\textrm{MeV}$. This is shown as ``CDD (H)" in yellow in Fig.~\ref{fig:graphic}.
The loop cross section has an accidental cancellation between different contributions, resulting in it being within the neutrino fog. 

A powerful probe of a TeV-scale Higgsino that does not depend on its relic abundance is the measurement of the electric dipole moment (EDM) of the electron. $CP$ violation is generically large in SUSY extensions of the Standard Model. As a result, charginos and neutralinos together induce an EDM on the electron. Once again, these vanish when the gaugino masses are very large. Currently, the EDM limits from ACME II~\cite{acme2018improved,Co:2021ion}, which constrain the EDM to $d_e < 1.1 \times 10^{-29}~\textrm{e~cm}$, rule out 10 TeV or heavier gauginos for $\tan \beta=2$, with limits degrading for larger $\tan \beta$. 
This corresponds to $\delta \approx 200 ~\text{MeV}$ Higgsino splitting. Projections from Advanced ACME, with sensitivity to $d_e \lesssim 10^{-30} \textrm{e~cm}$, corresponds to $100~\textrm{TeV}$ gauginos for $\tan \beta=2$, which can probe splittings down to 30 MeV. These are shown as the purple and blue blocks appropriately labeled in Fig.~\ref{fig:graphic}. The reach will be much weaker for larger $\tan \beta$. 

To summarize, the Higgsino splitting $\delta$, the only unknown parameter in the theory, is currently constrained  from above at the 100~MeV~--~GeV scale from EDM limits and Higgs exchange direct detection cross sections.  While these constraints are expected to improve, even in the most optimistic scenarios, future experiments are not expected to probe below $30~\textrm{MeV}$ splitting. 

This challenging window between $200~\textrm{keV}$ and $30~\textrm{MeV}$ has a remarkable property that might make new ideas for direct detection feasible. Suppose there is a process that successfully scatters $\tilde{H}_1$ to $\tilde{H}_2$. The $\tilde{H}_2$ is unstable, and for small enough mass splitting ($\delta < 1~\textrm{GeV}$), its dominant decay is to a $\tilde{H}_1$ and a photon. This process occurs radiatively via a W loop, and the resultant decay length for a Higgsino with typical DM velocity is given by~\cite{RadiativeDecay}
\begin{align} \label{eq: decay length}
    \lH=7.46~\text{km} \left(\frac{v}{400~\kms}\right) \left(\frac{400~\text{keV}}{\delta}\right)^3 ~, 
\end{align}
i.e., the mass splitting window that evades even optimistic projections from other methods corresponds to decay lengths well within terrestrial reach. This opens up a luminous dark matter direct detection prospect: if we can find a way to up-convert from the ground state to the excited state, we can observe the subsequent decay to a photon, which will produce a line signal. 

How do we increase sensitivity to larger splitting? The answer to this question is captured in~\Eq{\ref{eq:maxsplit}}. The existing limit is set by the heavy nuclei involved (typically Xe) and the maximum velocity of DM particles, currently assumed to follow the Standard Halo Model (SHM). Furthermore, as summarized in previous work~\cite{InelasticFrontier}, the existing limits do not reach the estimate in~\Eq{\ref{eq:maxsplit}} because inelastic scattering favors larger energy recoils, which are outside the low-background region targeted by existing experiments.

This work hinges on three key observations as applied to the luminous dark matter method. First, in the luminous DM process, the entire energy available in the center-of-momentum frame can be used to upscatter dark matter, unlike in collisional direct detection, where some energy has to go to the nuclear recoil to make it detectable. Second, in the luminous DM method, the nuclei act merely as springboards to up-scatter dark matter and are not involved in the detection itself. Therefore, heavy nuclei like Pb, or even U and Th, can be used, maximizing the reduced mass and thereby increasing sensitivity to the mass splitting. Finally, recent work has pointed to a dark matter subpopulation with origins in the LMC that populates the high-velocity tail of dark matter, exceeding the MW escape velocity. These points are explored in detail in the rest of the paper and correspond to the improvements seen in Fig.~\ref{fig:graphic} labeled ``this work."

\section{Luminous Detection}
In this section, we describe the process of luminous detection using large-volume neutrino detectors and calculate the expected signal rate based on a homogeneous approximation of the Earth.

In this paper, we use the term ``direct detection" to refer to any experiment that seeks to interact directly with relic DM particles.
Within the broad category of direct detection, we use the term ``collisional direct detection (CDD)" to refer to traditional WIMP detectors that search for nuclear recoils and ``luminous detection" for searches for their luminous signals.

\subsection{Terrestrial scattering and large-volume detectors}
The typical Higgsino decay length $\lH$ [\Eq{\ref{eq: decay length}}], which is on the order of a few kilometers, is much larger than the size of any detector but much smaller than the Earth's radius: $R_D \ll \lH \ll \REarth$. This hierarchy enables the luminous DM detection process:
\begin{enumerate}
    \item A Higgsino DM particle $\tilde{H}_1$ upscatters with a heavy nucleus in the Earth;
    \item The excited state $\Htwo$ subsequently travels a distance of $\sim \lH$ before decaying in a detector into $\Hone$ and a photon.
\end{enumerate}
In the rest frame of $\Htwo$, the signal photon energy is $E_{\gamma} = \frac{M_{\Htwo}^2 - M_{\Hone}^2}{2M_{\Htwo}} \approx \delta$ in the limit $\delta \ll M_{\Hone}$. Since the DM velocity is nonrelativistic, the lab frame observes negligible energy spread from Doppler shift. Hence, the resulting DM signature is a monoenergetic line of frequency $\delta$. Note that this process is not exclusive to Higgsinos, but applies to any inelastic DM model with a decay length that satisfies the aforementioned hierarchy. This mechanism was first explored in Ref.~\cite{LuminousDM} in the context of explaining the DAMA signal.

Despite the dual requirements that the excited particle must travel in the direction of the detector and decay within the detector, these geometrical constraints are precisely compensated by the large volume of the surrounding Earth. In fact, if the detector is located deep within the Earth, the Earth can be approximated as having infinite extent and homogeneous density, in which case the upscatterings and decays clearly form an equilibrium, occurring at the same rate within a given volume. Consequently, the luminous DM signal rate is roughly the same as the rate for direct collisions within the detector volume and is independent of the decay length. In fact, the rate depends solely on the detector volume and scales linearly with it. We will demonstrate this explicitly by calculating the signal rate using the homogeneous Earth approximation below.

Since the luminous signal scales with detector volume, a large-volume detector sensitive to photons in the hundreds of keV range is ideal. Reference~\cite{Luminous} noted that the underground neutrino detector Borexino~\cite{Borexino_Latest}, with a scintillator radius of $4.25$ meters and a threshold of \numcolor{$\Ethr = 150$ keV}, is well suited for this purpose. By treating both the radioactivity of the surrounding and the observed solar neutrinos as background, Ref.~\cite{Luminous} showed that an analysis of Borexino's existing data, looking for a daily modulation signal, can already set new limits on Higgsino DM. They also pointed out that the under-construction neutrino detector JUNO~\cite{JUNO_Conceptual}, with a radius of $17.5$ meters, has the potential to set even stronger limits in the future. Other large-volume neutrino detectors like Super-Kamiokande~\cite{superK}, Hyper-Kamiokande~\cite{HyperK}, and DUNE~\cite{DUNE} could be good detectors if their thresholds could be lowered.

We will describe several improvements on Ref.~\cite{Luminous} in later sections. Among other things, we will demonstrate that performing a line search instead of or in addition to a daily modulation search (see Sec.~\ref{sec: line search}) offers advantages. Additionally, we show that the SNO+ experiment~\cite{SNO+}, which has a radius of $6$ meters, can be added to the roster of large-volume neutrino detectors (see Table~\ref{tab: detectors}) with exciting potential for Higgsino discovery, despite its high latitude. We will also include KamLAND~\cite{Kamland} in our study.
Before discussing these and other new results, we first lay out the formalism used to compute the luminous DM signal rate.

\subsection{Signal rate calculation\label{sec:signal rate calculation}}
In the homogeneous Earth approximation, the detector is situated deep within the Earth, which is modeled as an infinite, uniform-density medium. Consider a single element with nuclear mass number $A$ present in the Earth; we will denote the nucleus or element by $A$. 
If a single nucleus has a Higgsino-nucleus scattering rate $\GxA$, then the total photon signal rate $\Gg$ can be obtained by multiplying the effective number of nuclei in the Earth that can produce photon signals
\begin{align}
    \Gg = \GxA n_A \Veff~,
\end{align}
where $n_A$ is the number density of element $A$ in the Earth and $\Veff$ is the effective volume contributing to the signal. In the homogeneous Earth approximation, each scattering event leads to a subsequent decay elsewhere in the Earth, resulting in a uniform distribution. Hence, the effective volume of target nuclei must be equal to the detector volume $V_D$, as we demonstrate below.

Let the center of the spherical detector with radius $R_D$ be at the origin of our coordinate system. Suppose at time $t=0$, an $\Ht$--$A$ scattering occurs outside the detector at position $\rbold$ (with $r>R_D$), where an $\Hone$ upscatters into an $\Htwo$. 
The fractional flux in the direction of the detector is determined by the fraction of the surface area of the spherical shell of radius $r$ centered at $\rbold$ that is covered by the circular projection of the detector, specifically $\dfrac{\pi R_D^2}{4\pi r^2}$. The probability of decay inside the detector is given by
\begin{align}
    P(r) = \int_{t_0(r)}^{t_f(r)} dt' \frac{1}{\tau} e^{-t'/\tau} \approx \frac{\left< d \right>}{\lH} e^{-r/\lH} ~,
\end{align}
where $\tau$ is the lifetime of $\Htwo$, $v$ is its velocity, the detector entrance time is $t_0(r) \approx r/v$, the detector exit time is $t_f(r) \approx \dfrac{r+\left< d \right>}{v}$, and we used that the decay length satisfies $\lH = v\tau \gg R_D$. The average distance the particle traverses in a straight line through a spherical detector can be shown to be $\left< d \right> = \dfrac{4}{3} R_D$. Therefore, the effective volume of the target nuclei that contribute to the signal is 
\begin{align} \label{eq:Veff estimate}
\Veff &\approx 4\pi \int_{R_D}^{\infty} dr ~ r^2 \left(\dfrac{\pi R_D^2}{4\pi r^2}\right) P(r) \\
&\approx \frac{4\pi}{3} R_D^3 \\
&= V_D ~,
\end{align}
which is equal to the volume of the spherical detector.

The scattering rate $\GxA$ is related to the Higgsino-nucleus cross section $\sxA$ by
\begin{align}
    \frac{d \GxA}{dE_R}= \nx \int_{v>\vmin(E_R)} d^3v~ \frac{d \sxA}{dE_R}~ v ~f^{\text{det}}(\vbold) ~,
\end{align}
where $\nx$ is the DM number density, $E_R$ is the nuclear recoil energy,  and $v$ is the DM velocity in Earth's frame. The function $f^{\text{det}}(\vbold)$ represents the time-averaged DM velocity distribution in the detector's frame. Since the experiment runs over many years, we take this time average. Although this procedure preserves the maximum velocity $\vmax$, it neglects the yearly modulation of the signal, a useful experimental handle~\cite{Luminous}.

The term $\vmin(E_R)$ is the minimum velocity required to produce a given recoil energy $E_R$ and mass splitting $\delta$. In the limits $\delta \ll \mx$, it is given by~\cite{Frontier}
\begin{align} \label{eq: vmin}
    \vmin(E_R) = \frac{1}{\sqrt{2 m_A E_R}} \left( \frac{m_A}{\muA} E_R + \delta \right) ~,
\end{align} 
where $\mu_A$ is the reduced mass relative to the nucleus and $m_A$ is the nucleus mass. Although $\vmin(E_R)$ depends on the recoil energy, there is also a global minimum velocity $\vglobalmin$, which is the minimum velocity required for any upscattering to occur. It is given by
\begin{align} \label{eq: vmin global}
    \vglobalmin = \sqrt{\frac{2 \delta}{\muA}} ~.
\end{align}

For a spin-independent interaction, the differential cross section is related to the total cross section $\sxA$ by
\begin{align}
    \frac{d\sxA}{dE_R}=\frac{m_A \sxA}{2\muA^2 v^2} ~.
\end{align}
Due to the coherent scattering of $\Ht$ with the entire nucleus, $\sxA$ is related to the Higgsino-nucleon cross section $\sxN$ by \cite{Barn}
\begin{align}
    \sxA = A^2 \left( \frac{\mu_A}{\mu_n} \right)^2 \sxN  ~F_A^2(E_R)\approx A^4 \sxN~F_A^2(E_R)~,
\end{align}
where $\mu_n \approx m_n \approx 1$ GeV is the reduced mass relative to a single nucleon. The breakdown of coherence is accounted for by the inclusion of the Helm form factor\footnote{Even at the largest mass splitting we consider, $\delta \sim 900$ keV, the dominant contribution to our integrand involving the form factor comes from momentum transfer around $q\sim 500$ MeV, which remains within the validity of the Helm form factor.}, which is given by~\cite{Luminous}
\begin{align} \label{eq: Helm form factor}
    F_A(E_R) = \frac{3}{q r_n} J_1(q r_n) e^{-q^2 s^2/2}~,
\end{align}
where $J_1(x)$ is a spherical Bessel function, $q=\sqrt{2m_A E_R}$ is the momentum transfer, $s \approx 0.9$~fm, and $r_n \approx 1.14 \left(\dfrac{A}{0.93}\right)^{1/3}$~fm. Combining these results, the differential scattering rate becomes
\begin{align}
    \frac{d \GxA}{dE_R}&= \nx \frac{m_A A^4 \sxN}{2\muA^2} ~\eta(\vmin(E_R)) ~F_A^2(E_R) ~,
\end{align}
where the mean inverse velocity, or halo integral $\eta$ (see Sec \ref{sec: halo integrals}) \cite{LMC}, is treated as a function of $E_R$:
\begin{align} \label{eq:halo integral 1}
    \eta(\vmin(E_R)) \equiv \int_{v>\vmin(E_R)} d^3v ~ \frac{1}{v}f^{\text{det}}(\vbold) ~.
\end{align}
Integrating over all nuclear recoil energies $E_R$, summing over all relevant elements in the Earth, and relating $\sxN$ back to the observable $\Gg$, we obtain
\begin{equation} \label{eq:final signal rate}
\begin{split}
    \Gg =& \sxN \times \Veff\nx  \sum_A n_A \frac{m_A A^4 }{2\muA^2}\\
    &\times\int_{0}^{\infty} dE_R \eta(\vmin(E_R)) F_A^2(E_R) ~.
\end{split}
\end{equation}
The DM number density is given by $\nx = \rho_{\rm DM}/\mx$, where we take $\rho_{\rm DM} = 0.3~\text{GeV}/\text{cm}^3$. We show how to use this equation to set 90\% confidence level limits in Sec.~\ref{sec:line search and south}.

Even though $\sxN=10^{-39} \textrm{cm}^2$ is fixed for the Higgsino, this equation applies to any inelastic dark matter with the same mass, $\mx=1.1$ TeV, and a decay length $l_{\text{DM}}$ satisfying $R_D \ll l_{\text{DM}} \ll \REarth$. Therefore, we will use the notation $\sigma_{\text{DM},n}$ for the general case. Note that when we consider adding supplemental Pb/U near the detector in Sec.~\ref{sec:supplement}, the signal rate will depend on the decay length, and those equations will only generalize to models with the same mass and decay length as the Higgsino.

It is important to note that this homogeneous Earth approximation neglects the directionality of the fastest DM population, which is particularly relevant for large mass splittings $\delta$. Within the SHM, the fact that the Sun moves through the MW implies that the fastest dark matter originates from the direction of the constellation Cygnus~\cite{Directional_DM, Luminous}. The effects of the LMC result in faster DM particles, but they share a similar directionality (see Sec.~\ref{sec:DM LMC}). This directionality induces a daily modulation in the luminous DM signal because, during the part of the sidereal day when Cygnus has risen above the Earth's horizon at the location of the detector, there may not be sufficient Earth volume for the Higgsino to scatter with on its way to the detector. Furthermore, since the Higgsino is much heavier than any nucleus, after scattering, it continues to move forward in Earth’s frame, with a small deflection angle of at most $\sim 10^{\circ}$~\cite{Luminous}. This forward scattering preserves the directionality of the DM signal. 

The primary impact of the inhomogeneity on our results is that the signal is suppressed for roughly half of each sidereal day, resulting in a signal time that is only half of the total runtime
\begin{align} \label{eq:signal time half}
    \tsignal = \ttotal/2 ~.
\end{align}
We explore this topic further in Sec.~\ref{sec: line search}. Nevertheless, aside from this adjustment, the effects of inhomogeneity average out, making the signal rate in the homogeneous approximation a reliable estimate. Moreover, the detector’s proximity to Earth’s surface does not lead to an overestimation of the signal, as the flux is predominantly unidirectional in reality.

\section{Dark Matter Populations\label{sec:DM populations}}
In this section, we introduce two halo models, the SHM and the LMC model. We also calculate the existing limits on inelastic DM from collisional direct detection using both halo models.
\subsection{Dark matter from the Large Magellanic Cloud\label{sec:DM LMC}}
Ref.~\cite{Luminous} treats the velocity distribution of dark matter using the SHM \cite{Original_SHM}, which, in the galactic frame, is given by a truncated Maxwell-Boltzmann distribution
\begin{align}
    f^{\text{gal}}_{\text{SHM}}(\vbold) ~ \propto ~ \exp\left(-\frac{v^2}{v_0^2}\right) \Theta(\vesc - v) ~,
\end{align}
where $v_0=220 ~\kms$ is the local standard of rest, \numcolor{$\vesc=540~\kms$} is the escape velocity of the MW, and $\Theta$ is the Heaviside step function. We have used the astrophysical parameters from Ref.~\cite{LMC}. We will take this SHM as one of our two example halo models. The SHM is the default model used in most analyses of direct detection experiments due to its simplicity and its natural explanation of flat rotation curves~\cite{Lisanti_TASI, SHM++}, despite its many known inaccuracies~\cite{SMH_Deviation_1, SMH_Deviation_2, SMH_Deviation_3, SMH_Deviation_4, SMH_Deviation_5}. However, the SHM is a poor approximation in the context of inelastic dark matter. Probing large mass splittings requires high DM velocities, but the high-velocity tail of the distribution is particularly sensitive to halo merger history~\cite{Lisanti_TASI, SMH_Deviation_4}. 

Recent studies \cite{LMCHighSpeed, LMC, LMC_Similar} argue that the high-velocity tail of the local dark matter distribution predominantly originates from the LMC and includes particles moving faster than the escape velocity of the MW. The LMC is a massive satellite galaxy that recently made its first close approach to the MW $\sim 50$ Myr ago~\cite{FirstInfall}. This encounter left behind two high-speed DM
populations in the solar neighborhood: particles originating from the LMC and MW particles that were accelerated by the LMC’s gravitational force. Since these two populations have roughly the same phenomenology, we will not distinguish between them and will collectively refer to them as ``dark matter from the LMC."

The effects of the LMC are prominent due to several factors. First, it has a large infall mass, approximately $10\%$ that of the MW,  thereby exerting a strong gravitational influence. Second, it only recently passed through the MW for the first time, so its particles have yet to equilibrate and can retain high velocities. Most importantly, the LMC happens to be moving in the opposite direction to the Sun~\cite{LMCHighSpeed}, leading to large relative velocities between the dark matter from the LMC and the Earth. This is a fortunate coincidence---had we lived on the other side of the galactic disc, the relative velocities would have been lower by $\sim 240~\kms$~\cite{LMCHighSpeed}.

The fact that the dark matter from the LMC is moving in the opposite direction from the Sun also has important implications for the daily modulation of the signal. In the SHM, the fastest dark matter comes from the direction of the Sun's motion with respect to the galaxy, which coincides with the direction of the constellation Cygnus~\cite{Directional_DM}. The proposal in Ref.~\cite{Luminous} relies on the fact that Cygnus is above the horizon for only part of the sidereal day for detectors at suitable latitudes, leading to a daily modulation in the signal. Since dark matter from the LMC also appears to originate from the direction of Cygnus, it would produce the same daily modulation signal. However, as we argue in Section \ref{sec: line search},  daily modulation is not essential for detecting a signal since we can perform a line search instead. Nevertheless, the daily modulation must be properly accounted for in our analysis and could also be a useful part of a real experimental analysis.

\begin{table*}[t!] 
    \centering
    \begin{tabular}{|c|c|c|c|c|c|c|c|c|} \hline 
         &  $M^{\text{MW}}_{\text{vir}} 
 ~[10^{11} \Msun]$ &$\rho_{\rm DM}~[\text{GeV}/\text{cm}^3]$&  $M^{\text{LMC}}_{\text{Infall}} ~[10^{11} \Msun]$ &  $r_{\text{peri}} ~[\text{kpc}]$ &$v_{\text{peri}}~[\kms]$& $t_{\text{peri}} ~[\text{Myr}]$&$r_{\text{pres}} ~[\text{kpc}]$ &$v_{\text{pres}} ~[\kms]$\\ \hline 
         observed&  10--20 &$\sim 0.3$&   $\sim$ 1--3 &  $\sim 48$  &$340 \pm 19$& $\sim -50$ & $\sim50$ &$321 \pm 24$\\ \hline 
         simulation&  12 &0.34&  3.2&  32.9 &$[205,376]$\footnote{This is the range of the pericenter velocities of all simulations (not just halo 13) in Ref.~\cite{LMC}. The specific value for halo 13 is not reported in that paper.}& -133 & 50.6&317\\ \hline
    \end{tabular}
    \caption{Comparison of the properties of the MW and LMC with their simulated analogs, referred to as halo 13 in Ref.~\cite{LMC}. $M^{\text{MW}}_{\text{vir}}$ is the virial mass of the MW~\cite{MWMass}. $\rho_{\rm DM}$ is the local dark matter density~\cite{DarkMatterDensity}. $M^{\text{LMC}}_{\text{Infall}}$ is the virial mass of LMC at infall. $r_{\text{peri}}$ and $v_{\text{peri}}$ are the distance~\cite{LMCHighSpeed} and velocity~\cite{LMC_velocity}, respectively, between the MW and LMC at closest approach (pericenter). $t_{\text{peri}}$ is the time of pericenter relative to the present day ($t=0$), and $r_{\text{pres}} $ and $v_{\text{pres}}$ are the present-day values.}
    \label{tab: halo 13}
\end{table*}

Reference~\cite{LMC} studies the LMC's effects using cosmological simulations that account for the formation history of the MW and the LMC. That paper utilized the Auriga cosmological magneto-hydrodynamical simulations~\cite{Auriga} and focused on a particular MW analog, referred to as halo 13 (Auriga ID Au-25), which has an LMC analog. The simulation was rerun with finer time resolution, and its present-day DM velocity distribution was computed. Table~\ref{tab: halo 13} shows that the MW's virial mass, local dark matter density, the LMC's infall mass, relative velocity at closest approach (pericenter), and present-day relative distance and velocity in the simulation are compatible with observed values. Although the distance and time of the closest approach deviate somewhat from the observed values, these parameters are realistic enough to be considered a reasonable proxy for the real MW-LMC system and were therefore used to analyze the results of direct detection experiments in that paper. We follow Ref.~\cite{LMC} in using the velocity distribution from this simulation (henceforth, the LMC model), acknowledging that it is not a perfect match to the real MW-LMC system. Future simulations may yield an even closer match. As we shall see, incorporating the effects of the LMC significantly boosts our sensitivity to Higgsinos and inelastic DM more generally. Possible effects of the Local Group or Local Supercluster were explored in Ref.~\cite{localGroup}.

\subsection{Halo integrals\label{sec: halo integrals}}
\begin{figure}[t!]
    \centering
    \includegraphics[width=0.5\textwidth]{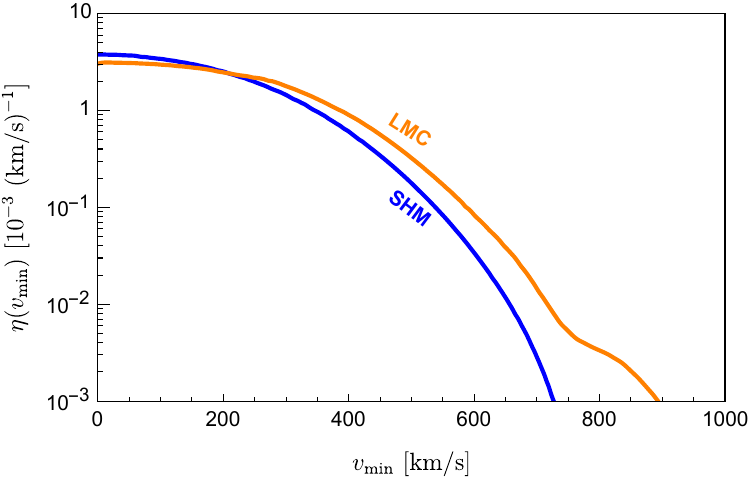}
    \caption{Time-averaged halo integrals of the SHM compared to the LMC model. The SHM is truncated at the MW escape velocity, \numcolor{$\vmaxSHM = 760~\kms$}, in the detector frame, while the LMC model reaches a higher maximum velocity \numcolor{$\vmaxLMC = 950~\kms$}. Figure adapted from Ref.~\cite{LMC}.}
    \label{fig: halo integral}
\end{figure}

A direct detection experiment's sensitivity to dark matter typically depends on the velocity distribution only indirectly, through the halo integral, which we now introduce, following Ref.~\cite{LMC}. Let $f^{\text{gal}} (\vbold)$ denote the present-day DM velocity distribution in the galactic frame. We transform this to the detector reference frame as
\begin{align}
f^{\text{det}}(\vbold,t) = f^{\text{gal}}(\vbold +\vbold_{\mathTerra}(t)+ \vbold_{\mathSun}) ~,
\end{align}
where $\vbold$ is the DM velocity in Earth's frame; $\vbold_{\mathTerra}(t)$ is Earth's velocity relative to the Sun, which varies over the course of a year; and $\vbold_{\mathSun}$ is the Sun's velocity relative to the galactic frame.  
For a given minimum DM lab-frame velocity $\vmin$ [see \Eq{\ref{eq: vmin}}], the mean inverse velocity above $\vmin$ is known as the halo integral \cite{LMC},
\begin{align}
    \eta(\vmin,t) \equiv \int_{v>\vmin} d^3v ~ \frac{1}{v} \fdet(\vbold,t) ~,
\end{align}
as introduced in \Eq{\ref{eq:halo integral 1}}.
We then take its time average,
\begin{align}
    \eta(\vmin) \equiv \langle \eta(\vmin,t) \rangle ~.
\end{align}
We extract $\eta(\vmin)$ for the LMC model from Ref.~\cite{LMC}, which is reproduced in Fig.~\ref{fig: halo integral}, along with that of the SHM for comparison. Since the SHM is truncated at the MW escape velocity \numcolor{$v^{\rm gal}_{\rm esc}=540~\kms$} in the galactic frame, this corresponds to a cutoff of \numcolor{$\vmaxSHM = 760~\kms$} in the detector frame. However, once the dark matter from the LMC is considered, the maximum DM velocity increases significantly to \numcolor{$\vmaxLMC = 950~\kms$}.

\subsection{Existing limits on inelastic dark matter}
Traditional WIMP detectors can detect inelastic dark matter scattering by measuring the nuclear recoil of the target nuclei, a process we refer to as collisional direct detection. As a new result of this paper, we calculate updated bounds on inelastic dark matter from collisional direct detection using both the SHM and the LMC model. Note that the mechanism of direct collisions with target nuclei is not specific to the Higgsino model, so the result is applicable to any inelastic dark matter with mass $\mx = 1.1$ TeV. However, the intersection of a constraint curve with the $\sigma_{\text{DM}, n}=10^{-39}~\text{cm}^2$ line delineates the corresponding limit on the Higgsino.

The collision event rate $\Gamma_{c}$ of a traditional WIMP detector is given by
\begin{align}
\begin{split}
    \Gamma_c =& \sigma_{\text{DM}, n} \times N_A n_{\text{DM}} \frac{m_A A^4 }{2\muA^2} \\
    &\times\int_{0}^{\infty} dE_R  \varepsilon(E_R) \eta(\vmin(E_R)) F_A^2(E_R) ~,
\end{split}
\end{align}
where $A$ is the nuclear mass number of the target nuclei [e.g., Xe or iodine (I)], $N_A$ is the number of target nuclei in the detector, $n_{\text{DM}}$ is the DM number density, and $\varepsilon(E_R)$ is the detection efficiency of the experiment. We use the halo integrals for the SHM\footnote{Note that other experiments, such as PandaX-II~\cite{PandaXII2022}, have claimed better sensitivity to Higgsinos than the numbers reported here because they used different versions of the SHM with different astrophysical parameters. We reanalyzed these experiments and compared them on the same footing using the same SHM. When we switched back to the specific SHM parameters used by the respective experiments, our calculations can mostly reproduce their results, with only small deviations.} and the LMC model found in Ref.~\cite{LMC}, reproduced in Fig.~\ref{fig: halo integral}.

\begin{figure}[htpb!]
    \centering
    \includegraphics[width=0.5\textwidth]{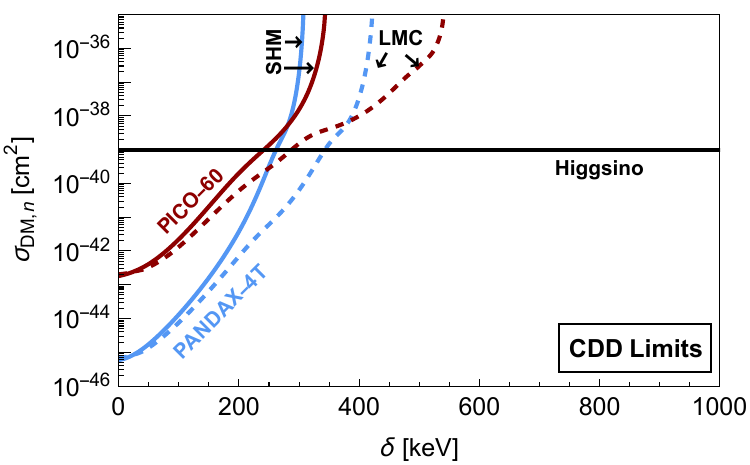}
    \caption{Our estimates for the current limits on inelastic dark matter from collisional direct detection (CDD), based on the PandaX-4T~\cite{PandaX4T2021} and PICO-60 $\text{CF}_3\text{I}$~\cite{PICO2023} experiments, are shown via the DM-nucleon cross section $\sigma_{\text{DM},n}$ vs splitting $\delta$.  The limits are shown for our two example halo models, the SHM and the LMC model.  The horizontal line shows the cross section for the standard Higgsino DM candidate.  These limits, derived from direct scattering inside the detectors, apply to any inelastic dark matter candidate---not just the Higgsino---provided the decay back to the ground state occurs outside the detector.}
    \label{fig:current limits}
\end{figure}

The results are shown in Fig.~\ref{fig:current limits}. The most stringent current limit on Higgsino is set by PandaX-4T~\cite{PandaX4T2021}, with \numcolor{$\delta^{\text{SHM}}_{\text{PandaX}} \sim 260$ keV} and \numcolor{$\delta^{\text{LMC}}_{\text{PandaX}} \sim 340$ keV}. These limits are also shown in Fig.~\ref{fig:graphic} as ``CDD (Z exchange)" in blue and ``+LMC" in orange, respectively. PICO-60 $\text{CF}_3\text{I}$~\cite{PICO2023} exhibits better sensitivity at higher $\delta$, but not for the Higgsino. Compared to other WIMP experiments, these two experiments benefit from higher efficiencies at large recoil energies, with sensitivity to recoil energies up to $\sim 120$ keV and $\sim $ MeV, respectively.

Some WIMP experiments have collected high recoil-energy data but excluded them from their analyses, resulting in low efficiencies at high recoil energy and inadvertently limiting their sensitivities to inelastic DM at high $\delta$. As recommended in previous works~\cite{InelasticFrontier}, we advocate reanalyzing existing data over a larger range of recoil energy, as this presents an easy opportunity to improve current inelastic DM bounds. 

\section{Terrestrial Heavy Elements\label{sec:heavy elements}}
The maximum $\delta$ that can be upscattered by the element $A$ is determined using \Eq{\ref{eq: vmin global}},
\begin{align} \label{eq: delta max A approx}
    \deltamaxA =\frac{1}{2} \mu_A \vmax^2 \approx \frac{1}{2} A m_n \vmax^2 ~,
\end{align}
where $m_n \approx 1$ GeV is the nucleon mass, and $\vmax$ is the maximum lab-frame DM velocity. In our two halo models, \numcolor{$\vmaxSHM=760~\kms$} and \numcolor{$\vmaxLMC = 950~\kms$}. Given our limited sensitivity to larger Higgsino mass splittings $\delta$, it becomes essential to account for any heavy elements in the Earth with appreciable abundances. Reference~\cite{Luminous} focused on the trace abundances of Pb ($A=207$).

\begin{figure}[htpb!]
    \includegraphics[width=0.5\textwidth]{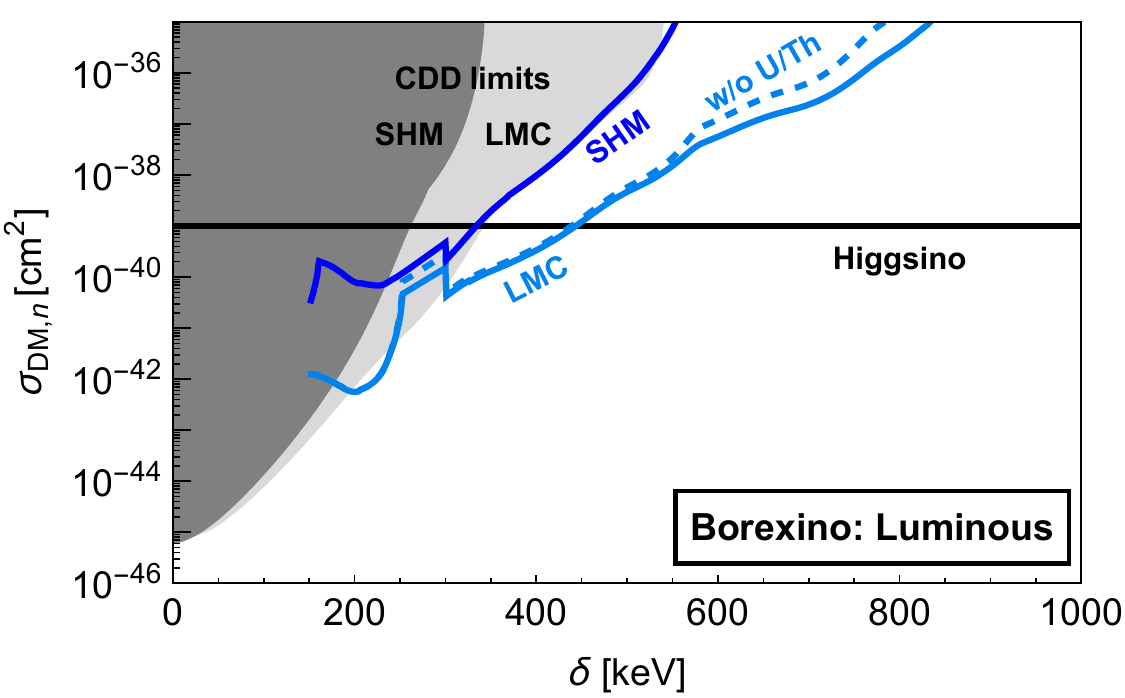}
    \caption{We show projections for the sensitivity of the Borexino experiment~\cite{Borexino_Latest} to Higgsino dark matter using the luminous detection strategy, if Borexino were to analyze their existing data to search for this signal.  Here, $\sigma_{\text{DM},n}$ is the DM-nucleon cross section, and $\delta$ is the mass splitting. These projections apply to any inelastic dark matter with a mass of $1.1~\text{TeV}$ and decay length much larger than the detector but much smaller than the Earth. The horizontal line indicates the actual cross section that Higgsinos would have in the minimal model.  The projections assume our two different example halo models, the SHM and LMC.  We have also illustrated the effect on the LMC curve when the uranium and thorium content of the Earth is neglected.  The gray bands show the existing limits from collisional direct detection (CDD) experiments as in Figure \ref{fig:current limits}, assuming the SHM (dark gray) or LMC (light gray) halo models. }
    \label{fig:Borexino}
\end{figure}

In this paper, we consider two heavier elements, Th ($A=232$) and U ($A=238$), whose number densities in Earth's crust are smaller than that of Pb by factors of 2 and 10, respectively; see Table~\ref{tab: elements}. Their lower abundances are partially compensated by the $A^4$ scaling in \Eq{\ref{eq:final signal rate}}, which arises from the coherent scattering of the Higgsino with the entire nucleus~\cite{Barn}. The significance of Th and U is amplified by the much larger $(\vmaxLMC)^2$ term in \Eq{\ref{eq: delta max A approx}}. As shown in Table~\ref{tab: elements}, U can upscatter a maximum mass splitting of \numcolor{$\deltamaxLMC{U} \sim 940$ keV}, exceeding Pb's reach by over $100$~keV. Although this 100-keV improvement is most readily applicable to larger cross sections and is not immediately realized on the Higgsino line in the simplest searches (see Fig.~\ref{fig:Borexino}), more aggressive strategies---such as adding U/Pb supplements around the detector, as discussed in Sec.~\ref{sec:supplement}---can be employed to lower the constraint curves (see Figs.~\ref{fig:overall SNO} and \ref{fig:overall JUNO}), making Th and U relevant for the Higgsino. Furthermore, beyond the Higgsino, more general inelastic DM models with larger cross sections would more readily benefit from this improvement.

\begin{table*}[htpb!]
    \centering
    \begin{tabular}{|c|c|c|c|c|c|c|c|c|} \hline 
         element&   $A$&$n_A~[\text{km}^{-3}]$&  $a_A$& relevant source& $\deltamaxALMC~[\text{keV}]$&$\lmaxALMC~[\text{km}]$ & $\deltamaxASHM~[\text{keV}]$&$\lmaxASHM~[\text{km}]$\\ \hline 
 Ca& 40.078& $1.8 \times 10^{36}$& 0.0261& mantle& 180& $ 180$& 120&580\\\hline 
         Fe&   55.85&$3.0 \times 10^{36}$&  0.063&  mantle& $ 250$&$71$& 160&220\\ \hline 
 \multirow{2}{*}{Sr}&  \multirow{2}{*}{87.62}& $6.8 \times 10^{32}$& $22 \times 10^{-6}$& mantle&  \multirow{2}{*}{$390$}&  \multirow{2}{*}{$ 20$}&  \multirow{2}{*}{240}&  \multirow{2}{*}{62}\\ \cline{3-5} 
 & & $6.2 \times 10^{33}$& $320 \times 10^{-6}$& total crust& & & &\\ \hline 
  \multirow{2}{*}{Ba} &  \multirow{2}{*}{137.33} & $1.4 \times 10^{32}$& $6.85 \times 10^{-6}$& mantle&  \multirow{2}{*}{$ 580$}&  \multirow{2}{*}{$ 5.8$}&  \multirow{2}{*}{370}&  \multirow{2}{*}{18}\\ \cline{3-5}
 & & $5.7 \times 10^{33}$& $456 \times 10^{-6}$& total crust& & & &\\ \hline
         Pb&   207.2&$1.4 \times 10^{32}$&  $17 \times 10^{-6}$& upper crust & $ 830$&$2.0$& 530&6.1\\\hline 
         Th&   232.04&$7.7 \times 10^{31}$&  $10.5 \times 10^{-6}$& upper crust & $ 920$&$1.5$& 580&4.6\\ \hline 
         U&   238.03&$1.9 \times 10^{31}$&  $2.7 \times 10^{-6}$& upper crust & $ 940$&$ 1.4$& 590&4.3\\ \hline
    \end{tabular}
\caption{Relevant elements in the Earth. Listed are the nuclear mass number $A$, their number density $n_A$, their mass abundances $a_A$ in the crust~\cite{Composition_Crust} and mantle~\cite{Mantle_Composition}. We also show the maximum mass splitting each element can scatter and the corresponding decay length for the two example halo models, the SHM and LMC model. See Appendix~\ref{sec:terrestrial elements} for further details.}
    \label{tab: elements}
\end{table*}
In Table~\ref{tab: elements}, we also show other terrestrial elements considered in addition to the heavy elements Pb, Th, and U. 
Among the major elements in the mantle, only calcium (Ca) and iron (Fe) can scatter with splittings $\deltamaxALMC$ above Borexino's threshold \numcolor{$\Ethr=150$ keV}. We also include barium (Ba) and strontium (Sr), which have intermediate values of $\deltamaxA$ and appreciable abundances in the mantle and crust. Additionally, we list the Higgsino decay lengths at $\deltamaxA$ for each element, given by
\begin{align}\label{eq:lmaxA}
    \lmaxA \equiv \lH (\vmax,\deltamaxA) ~,
\end{align}
for both the SHM and LMC model.
While this is not the maximum distance over which an element can have an impact---since it can always scatter at lower $\delta$, which would travel farther---it serves as a useful length scale associated with each element. The details of computing number densities and how we incorporate the varying densities in Earth’s crust and mantle are provided in Appendix~\ref{sec:terrestrial elements}.

\section{Advanced Luminous Detection Techniques\label{sec:advanced detection}}
In this section, we introduce three strategies to enhance luminous detection of Higgsino dark matter. Specifically, in the following three subsections, we propose a method to enhance signals, a preliminary idea to reduce background, and a new analysis approach that make more detectors viable for this purpose.

\subsection{Uranium and lead supplements\label{sec:supplement}}
As previously explained, new limits on Higgsino DM can be set by reanalyzing existing data from Borexino, as well as by performing similar analyses with upcoming detectors such as JUNO, SNO+, and KamLAND. The most stringent limit of this type can be achieved by JUNO due to its larger volume. However, as shown in Fig.~\ref{fig:Juno LMC}, although it is kinematically possible for U to excite mass splittings up to \numcolor{$\delta^{\text{LMC}}_{\text{max,U}} \sim 940~\text{keV}$}, JUNO can only set a limit of \numcolor{$\delta^{\text{LMC}}_{\text{JUNO}} \sim 580$} keV with a 10-year runtime. While this represents a significant improvement over the current limit of \numcolor{$\delta^{\text{LMC}}_{\text{PandaX}} \sim 340$ keV}, it still falls short of its kinematic potential. Further sensitivity will be necessary to lower the constraint curves.

To this end, we propose installing a large volume of Pb or U near the detector, which would dramatically increase the number density of these heavy elements compared to their terrestrial trace abundances. Furthermore, the scattering rate with the installed volume does not exhibit daily modulation, effectively doubling its exposure time relative to terrestrial elements. Note that the line search method is required to detect such a signal (see Sec.~\ref{sec: line search}). Another advantage of the Pb/U supplement is its flexibility, as it can be moved or removed to further reduce background or confirm any potential signals, although doing so would forfeit the factor of 2 in exposure time. Acquiring a large volume of either element appears feasible in terms of cost: Pb is relatively inexpensive at \$2 USD per kg, and there are $\sim 10^6$ tons of depleted U worldwide, with limited use~\cite{depleteduranium}.

Although the Pb/U supplement provides a much larger number density $\npb \gg n_{\text{pb,Earth}}$ in the signal rate formula \Eq{\ref{eq:final signal rate}}, the effective volume term $\Veff$, which previously represented the relevant volume of the surrounding Earth, is now replaced by the much smaller effective volume of the Pb/U material. We calculate this modified effective volume below using a toy model. While we refer to Pb for simplicity, the result is identical for U.

In our toy model, we assume the Pb supplement is distributed into a spherical shell with an outer radius $\Rpb$ surrounding the spherical detector with radius $R_D$. In practice, the Pb would optimally be placed as close to the detector as possible. 
The effective volume $\Veff$, calculated in \Eq{\ref{eq:Veff estimate}}, which is now contributed by the Pb supplement rather than the surrounding Earth, is suppressed by the decay length $\lH$
\begin{align} 
\Veffpb &\approx 4\pi \int_{R_D}^{\Rpb} dr ~ r^2 \left(\dfrac{\pi R_D^2}{4\pi r^2}\right) \frac{4R_D/3}{\lH} e^{-r/\lH} \\
&\approx V_D \left(\frac{\Rpb-R_D}{\lH} \right)~.
\end{align}
We take the velocity-dependence of $\lH$ to be the minimum velocity $\vglobalmin$ [\eq{\ref{eq: vmin global}}] required to upscatter against Pb.
Expressing the effective volume of Pb in terms of the actual volume of the Pb supplement $\Vpb = \frac{4\pi}{3}(\Rpb^3 - R_D^3)$, we obtain
\begin{align} \label{eq:Veff pb}
\Veffpb 
&\approx V_D \left(\frac{1}{\lH} \right) \left[\left(\frac{3}{4\pi}\Vpb + R_D^3 \right)^{1/3}-R_D \right]~.
\end{align}

\begin{figure}[htbp!]
    \centering
    \begin{subfigure}{0.5\textwidth}
        \centering
        \includegraphics[width=\textwidth]{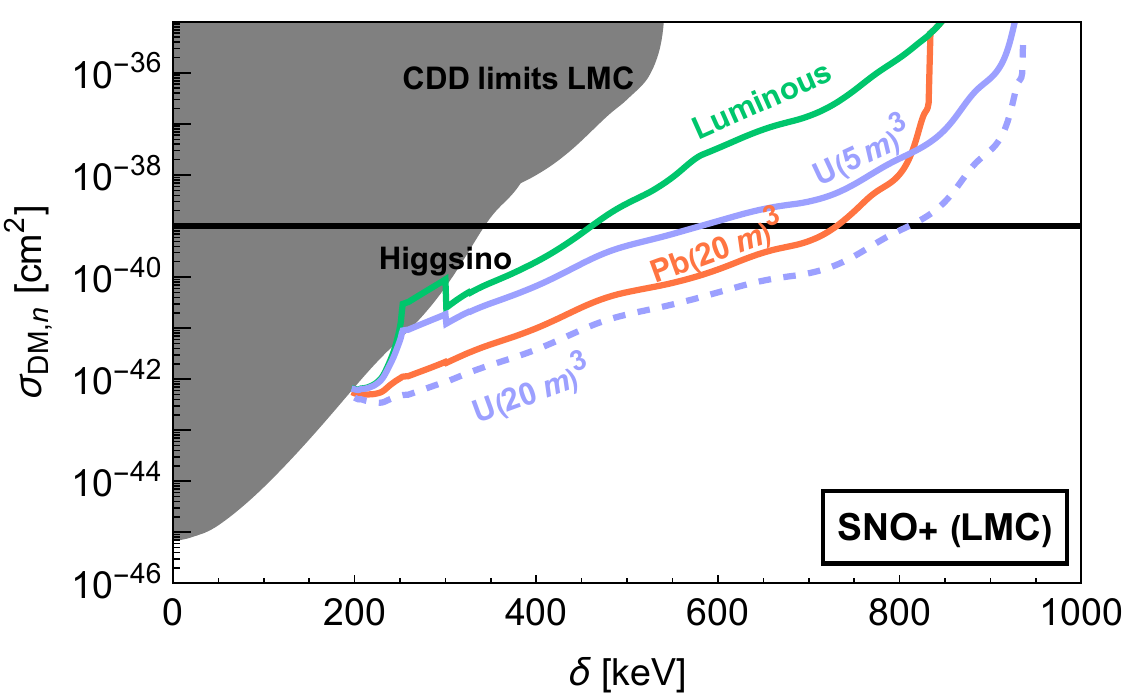}
        \caption{}
        \label{fig:sno LMC}
    \end{subfigure}
    
    \vspace{1cm}
    
    \begin{subfigure}{0.5\textwidth}
        \centering
        \includegraphics[width=\textwidth]{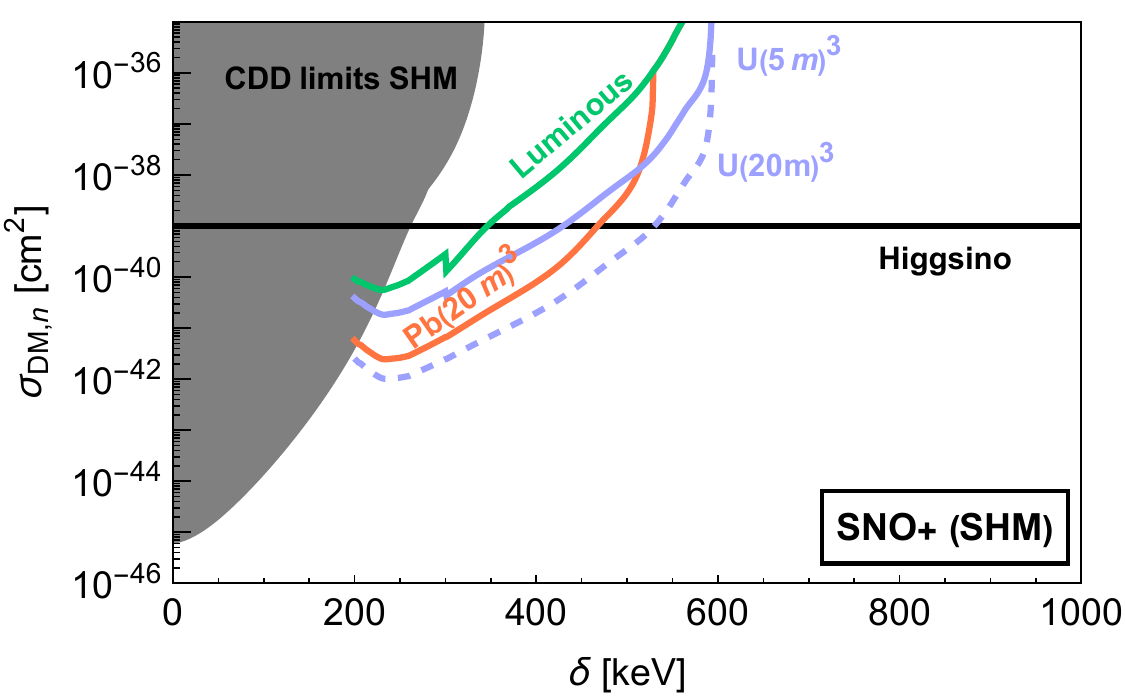}
        \caption{}
        \label{fig:sno SHM}
    \end{subfigure}
    
    \caption{Projected sensitivity for the SNO+ experiment~\cite{SNO+Threshold} in the DM-nucleon cross section $\sigma_{\text{DM},n}$ vs splitting $\delta$, if the luminous detection analysis discussed here is carried out.  We also show projections if extra heavy material is placed near SNO+ using as example scenarios: $(5 \, \text{m})^3$ of uranium and $(20 \, \text{m})^3$ of lead or uranium.  The two different figures show sensitivities assuming our two different example halo models (a) LMC and (b) SHM. The luminous projection in green applies to any inelastic dark matter with a mass of $1.1~\text{TeV}$ and decay length much larger than the detector but much smaller than the Earth. The projections with supplemental Pb/U  also assume that the decay length is the same as that of the Higgsino. The horizontal line shows the actual cross section that Higgsinos would have in the minimal model.  Shown in gray are current limits from collisional direct detection (CDD) experiments as in Fig.~\ref{fig:current limits}.}
    \label{fig:overall SNO}
\end{figure}

If $\Vpb$ is much smaller than the detector volume, the effective volume increases proportionally with $\Vpb$,
\begin{align}
\Veffpb \approx V_D \times \frac{1}{4\pi}\frac{\Vpb}{\lH R_D^2} ~.
\end{align}
However, once $\Vpb$ becomes much larger than the detector volume, the effective volume increases only linearly with $\Rpb$. Therefore, it is most cost-effective to install a Pb/U supplement with a volume comparable to the detector volume.  Since JUNO has a larger volume than SNO+ and KamLAND, it continues to benefit for longer as more Pb/U is added.

\begin{figure}[t!]
    \centering
    \begin{subfigure}{0.5\textwidth}
        \centering
        \includegraphics[width=\textwidth]{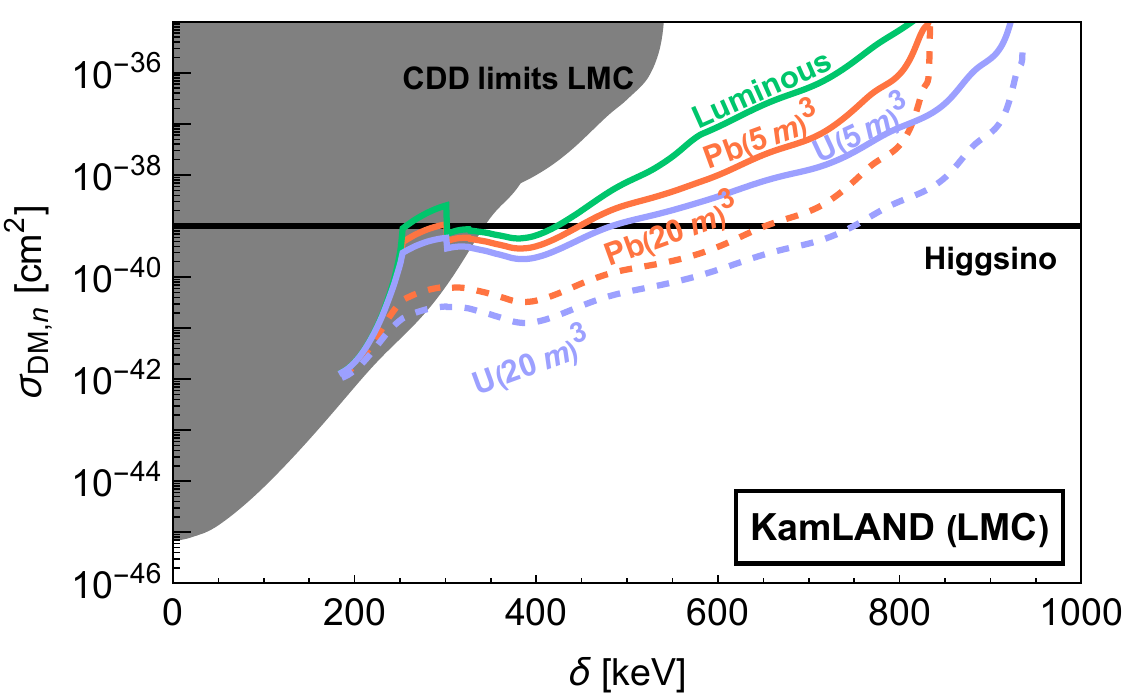}
        \caption{}
        \label{fig:Kamland LMC}
    \end{subfigure}
    
    \vspace{1cm}
    
    \begin{subfigure}{0.5\textwidth}
        \centering
        \includegraphics[width=\textwidth]{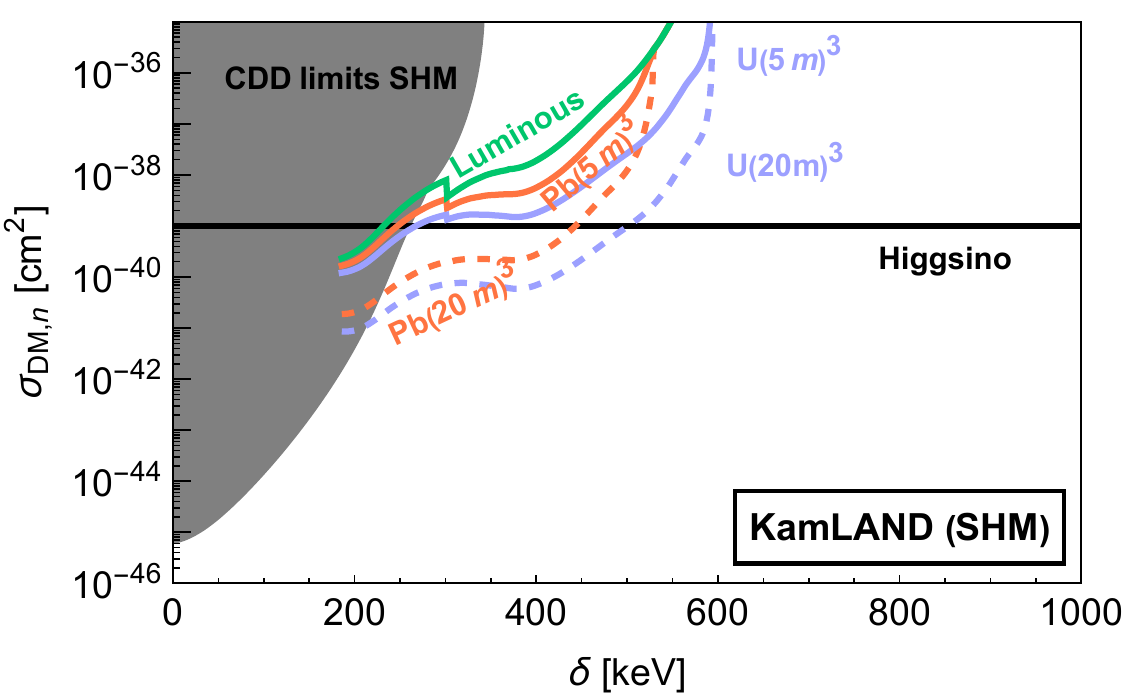}
        \caption{}
        \label{fig:Kamland SHM}
    \end{subfigure}
    
    \caption{Same as Fig. \ref{fig:overall SNO} except for the KamLAND experiment~\cite{Kamland} instead of SNO+.}
    \label{fig:overall Kamland}
\end{figure}

\begin{figure}[t!]
    \centering
    \begin{subfigure}{0.5\textwidth}
        \centering
        \includegraphics[width=\textwidth]{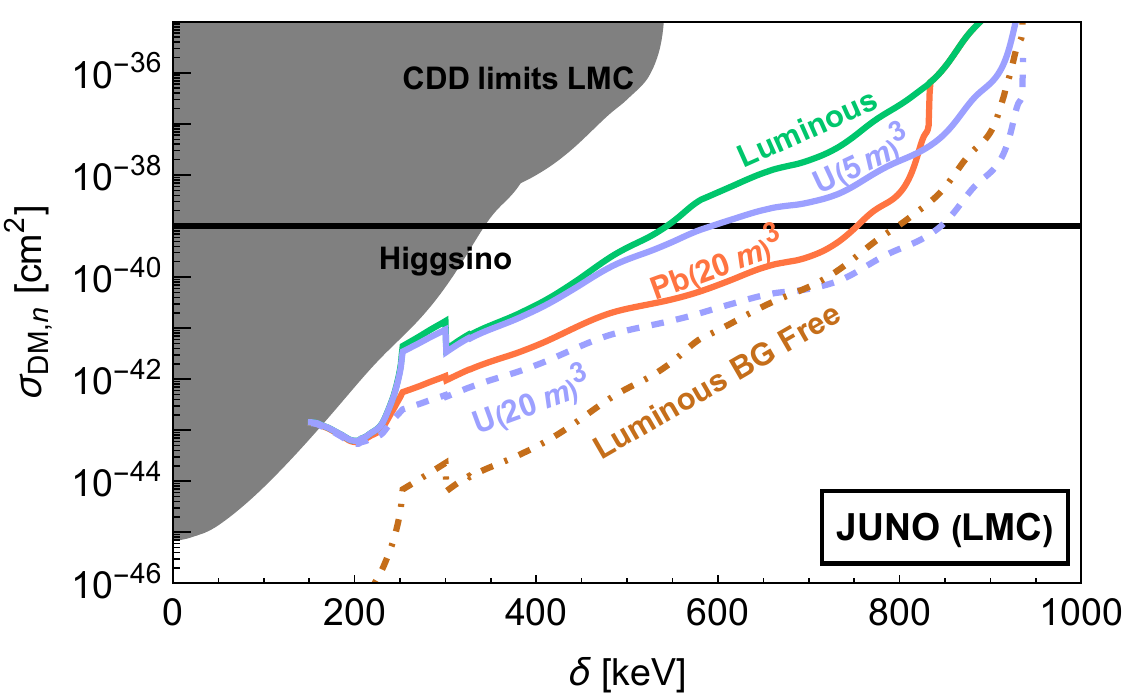}
        \caption{}
        \label{fig:Juno LMC}
    \end{subfigure}
    
    \vspace{1cm} 
    
    \begin{subfigure}{0.5\textwidth}
        \centering
        \includegraphics[width=\textwidth]{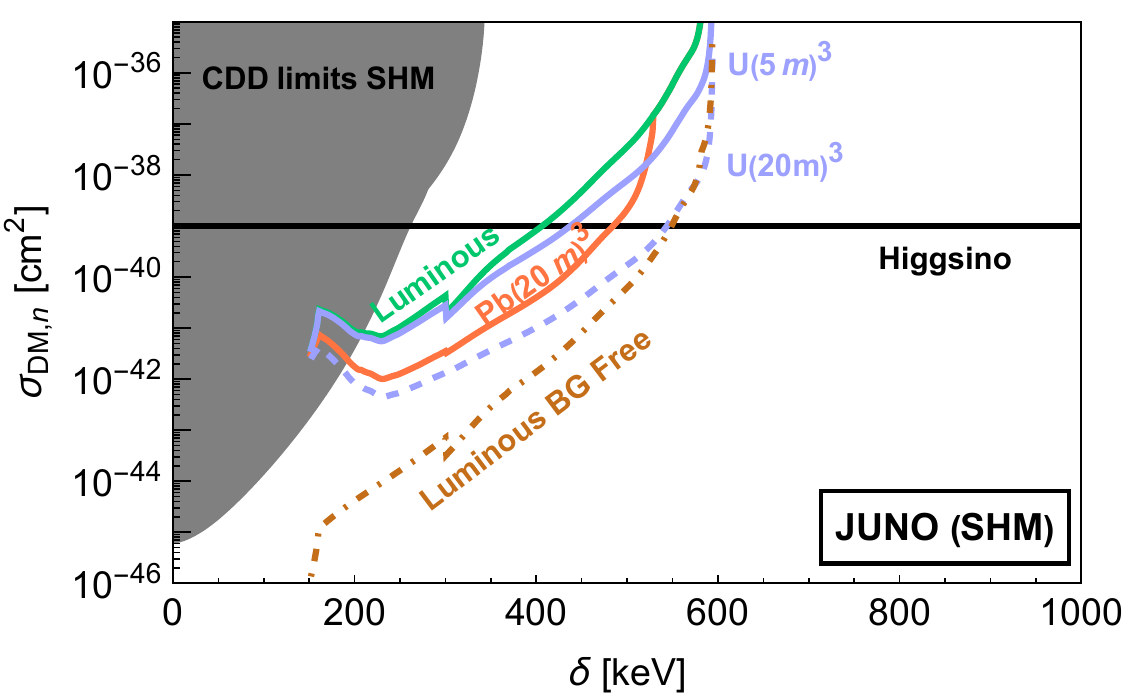}
        \caption{}
        \label{fig:JUNO SHM}
    \end{subfigure}
    
    \caption{Same as Fig. \ref{fig:overall SNO} except for the JUNO experiment~\cite{JUNO_Conceptual} instead of SNO+.  Here, we also plot a curve assuming a background-free detector the same size as JUNO (see Sec.~\ref{sec:BGF}). While complete elimination of backgrounds may never be fully achievable in practice, this curve demonstrates the limit of background reduction.}
    \label{fig:overall JUNO}
\end{figure}

Another important insight from \Eq{\ref{eq:Veff pb}} is that since the decay length $\lH$ is inversely proportional to $\delta^3$ [\Eq{\ref{eq: decay length}}], there is much more severe decay length suppression at low $\delta$, which improves rapidly as $\delta$ increases. At sufficiently low $\delta$, the Pb/U supplement would provide little to no advantage. Fortunately, as shown in Figs.~\ref{fig:overall SNO}(\subref{fig:sno LMC}), \ref{fig:overall Kamland}(\subref{fig:Kamland LMC}), and \ref{fig:overall JUNO}(\subref{fig:Juno LMC}), the Higgsino line crosses the constraint curves at a $\delta$ that benefits significantly from the Pb/U supplement for all three experiments.

Various strategies are available to prevent the supplemental Pb/U from inducing additional radiation background in the neutrino detectors. First, these experiments already use position reconstruction to apply fiducial volume cuts, which can eliminate the vast majority of external $\gamma$-ray background (external $\alpha$'s and $\beta$'s are not capable of reaching the inner volume)~\cite{Borexino_Phase_I,JUNO_Fiducial_Cut, SNO_Fiducial_Cut}. Second, these detectors are submerged in large water pools several meters thick, providing substantial radiation shielding. If the volume of the supplemental material is smaller than that of the water pool, reducing the water volume could help keep the new material closer to the detector. However, if the Pb/U volume is significantly larger, this consideration becomes irrelevant. In scenarios where the $\gamma$-radiation is the dominant background, we estimate that adding a 60~cm-thick layer of archaeological lead~\cite{archaeological_lead} shielding between the supplemental Pb/U and the detector would reduce $\gamma$-radiation by at least a factor of $\sim 10^6$, bringing it below the level produced by their terrestrial abundances~\cite{gamma_shielding}. Additionally, potential neutron activation due to $(\alpha,n)$ reactions can be shielded by $\mathcal{O}(10~\text{cm})$ of water~\cite{neutron_shielding}. The exact implementation of these radiation shielding strategies requires further research and input from members of these collaborations.

\subsection{Towards background-free detection\label{sec:BGF}}

Aside from being limited by their volumes, the primary obstacle preventing the neutrino detectors from achieving better sensitivity is background, which we have defined as including both the radioactivity of the surroundings as well as observed solar neutrinos. While it is possible to distinguish solar neutrinos by filtering out signals originating from the direction of the Sun, reducing background from radioactivity is much more difficult due to irreducible radioactive impurities. Ref.~\cite{Luminous} considered the sensitivity of gaseous scintillation detectors such as CYGNUS~\cite{CYGNUS}, which experience less background than liquid scintillators due to their lower density. However, CYGNUS is only sensitive to lower values of $\delta$, whereas our focus is on higher $\delta$ for the Higgsino.

Here, we discuss a preliminary idea for a dedicated, background-free $\gamma$-ray detector for Higgsino DM. Since the standard luminous detection scheme at JUNO can already probe up to $\sim 600$ keV [see Fig.~\ref{fig:overall JUNO}(\subref{fig:Juno LMC})], we focus on the backgrounds above this energy. Assuming complete rejection of the solar neutrino background is achievable, the remaining backgrounds between $600$ keV and $1$ MeV arise from $\beta$-decays of ${}^{210}\text{Bi}$, ${}^{85}\text{Kr}$, ${}^{40}\text{K}$, and ${}^{11}\text{C}$, as well as $\alpha$-decays of ${}^{238}\text{U}$ and ${}^{232}\text{Th}$ (see Fig.~1.10 of Ref.~\cite{JUNO_Conceptual}). Crucially, all of these backgrounds are generated by charged particles, while our signal is produced by photons. Therefore, complete background rejection may be possible if we can reliably distinguish between photons and charged particles.

Despite the challenge of achieving this below a few MeV due to the unavailability of pair production, there have been proposals for detectors of this kind in an astrophysics context, such as ASTROGAM, whose Anticoincidence detector has a projected background rejection efficiency of over $99.99\%$~\cite{Astrogam}. 
While complete elimination of backgrounds may never be fully achievable in practice, we present background-free projections in Fig.~\ref{fig:overall JUNO} as an ultimate limit for such a dedicated detector.

We are primarily considering the background-free detector as an alternative to the simpler but potentially more costly approach of adding a Pb/U supplement, which also targets the same energy range.  It is also possible to combine these two ideas, which will be briefly discussed in Sec.~\ref{sec:projection}.

\subsection{Line searches and detector depths\label{sec: line search}}
\subsubsection{Cygnus and daily modulation}
As mentioned previously, due to the Sun’s motion relative to the galaxy, most DM particles, especially the high-speed ones, appear to originate from the direction of the constellation Cygnus in Earth’s frame~\cite{Directional_DM, Luminous}. This effect is well-known within the SHM  and remains valid even after accounting for the effects of the LMC, as explained in Sec.~\ref{sec:DM LMC}. Moreover, since the Higgsino is much heavier than any target nucleus, the excited Higgsino continues to travel forward after upscattering, with a small deflection angle of at most $\sim 10^{\circ}$, allowing the luminous signal to retain its Cygnus-directionality~\cite{Luminous}.

Since Cygnus has a declination of $45^\circ$~North, detectors with latitudes above $45^\circ$~North always see Cygnus above the horizon, while for detectors below $45^\circ$~South, Cygnus is always below the horizon. Detectors located in the intermediate region between these two latitudes see Cygnus above the horizon only for some fractions of the day. Reference~\cite{Luminous} argues that only detectors in this intermediate region can be used to search for inelastic DM effectively for the following reason: while the high-speed DM particle has to pass through the bulk of the Earth to reach the detector when Cygnus is below the horizon, it passes through a much smaller fraction of the Earth when Cygnus is above the horizon. This significantly reduces its signal since there are fewer materials to scatter with, introducing a sidereal daily modulation to the signal. A limit can then be set by requiring that the signal when Cygnus is below the horizon is no larger than the statistical fluctuation of the background, predicted by the measured values when Cygnus is above the horizon.

We note two important additional considerations. First, in addition to a daily modulation search, a line search could also be considered. This approach makes potential detectors located below $45^\circ$ South viable candidates. Second, for detectors located above $45^\circ$ North, the depths of their underground locations should be carefully considered. Upon accounting for this, certain detectors, notably SNO+, become viable for Higgsino dark matter detection. We demonstrate these two points in the following subsections.

\subsubsection{Line searches and high southern latitudes\label{sec:line search and south}}
\begin{table*}[t!]
    \centering
    \begin{tabular}{|c|c|c|c|c|c|c|c|} \hline 
         detector&  radius [m] &mass [t]&  depth [km]& latitude  &$\hmax$ [km] & $\Ethr$ [keV] &$\deltaminFe$ [keV] \\ \hline 
         Borexino&  4.25 &278&  1.4& $42.3^{\circ}$ North&\numcolor{630}&150 &\numcolor{110}\\ \hline 
         JUNO&  17.5 &20000&  0.45& $22.1^{\circ}$ North  &\numcolor{5000}&150\footnote{Threshold assumed to be the same as Borexino.}&46 \\ \hline 
         SNO+&  6 &780&  2.07 & $46.5^{\circ}$ North  &66 & $\sim 200-400$&\numcolor{260}\\\hline
 KamLAND& 6.5& 1000& $\sim 1$& $36.4^{\circ}$ North& 1900& 200&70\\\hline
    \end{tabular}
    \caption{Large underground neutrino detectors, their radii, masses, depths~\cite{JUNO_Depth, SNO+Depth, KamlandDepth}, latitudes, maximum Earth-penetration lengths, experimental thresholds, and minimum $\delta$ effectively probed by iron.}
    \label{tab: detectors}
\end{table*}
While daily modulation is an important effect and could play a useful role in a real experimental analysis, it is not essential for detecting a signal. Instead, we can perform a line search: by dividing the data into frequency bins based on the experiment’s frequency resolution, we search for a potential signal in any bin that exceeds the statistical fluctuations of the neighboring bins. These neighboring bins can be treated as background, as the dark matter decay signal appears as a distinct line at the frequency $\delta$. Note that both neutrinos and radioactive decay are considered as background in this context.

Concretely, our treatment of the background is as follows. Let the background signal rate be denoted as $\Gammabackground$. We extract the mass-normalized background spectrum for Borexino from Figure 6 of Ref.~\cite{Luminous} and assume that other detectors will have similar background rates per scintillator mass. Note that we have assumed the photon signal is equivalent to that produced by an electron with the same energy, neglecting the effects of quenching, which can cause a $\sim 10\%$ shift in the spectrum~\cite{Borexino_Electron_Decay}. We integrate the background spectrum to obtain $\Gammabackground$, accounting for Borexino’s approximately 10\% energy resolution (e.g., $\pm 0.1 \delta$ around a given $\delta$)~\cite{Borexino_Electron_Decay}. We assume SNO+ has the same resolution, while JUNO offers a better energy resolution of $\pm \dfrac{0.03~\delta}{\sqrt{\delta/\text{1 MeV}}} $~\cite{JUNO_Conceptual}. For KamLAND, we used their published background~\cite{Kamland} and the energy resolution of $\pm \dfrac{0.04~\delta}{\sqrt{\delta/\text{1 MeV}}} $ of KamLAND2-Zen~\cite{ChristGrant}.

Let the total run time be denoted as $\ttotal$. We will use Borexino's total run time of $\ttotal=3628.7~\text{days}\approx 10~\text{years}$ and assume similar run times will be possible for other upcoming detectors. Due to the daily modulation of the signal, we approximate that the signal is present for half of each sidereal day, yielding a signal time of $\tsignal = \ttotal/2$. The total number of background photons is given by $\Gammabackground \tsignal$.

To set a 90\% confidence level, we perform a one-sided Gaussian integral, requiring the signal to be less than $1.28$ times the Poisson standard deviation of the background
\begin{align}
    \Gg = \Gammasignal \leq 1.28 \sqrt{\frac{\Gammabackground}{\tsignal}} ~.
\end{align}
In the case of a background-free detector (see Sec.~\ref{sec:BGF}), we take the 90\% confidence level of a null experiment with Poisson distribution~\cite{StatsBook}
\begin{align}
    \Gg = \Gammasignal \leq -\frac{1}{\tsignal} \log(1-0.9) ~.
\end{align}
Substituting these estimates of $\Gg$ into \Eq{\ref{eq:final signal rate}} and numerically evaluating the integral allows us to set 90\% confidence-level limits on the DM-nucleon cross section $\sigma_{\text{DM},n}$ as a function of $\delta$.

In the scenario where supplemental Pb/U is installed next to the detector, as discussed in Sec.~\ref{sec:supplement}, the additional Pb/U scatters at a constant rate and does not experience daily modulation. Therefore, their signal must be identified using line searches, and their exposure time is doubled compared to the normal scenario.

Since this line search method does not require any ``signal-off time" to measure the background, potential detectors located below $45^{\circ}$ South are excellent candidates for our purpose. In this region, DM passes through the bulk of the Earth throughout the entire sidereal day, maximizing the integration time. Although no neutrino detectors currently exist in this area, it is important not to exclude potential future detectors. Notably, the proposed 3-kiloton solar neutrino detector at ANDES~\cite{ANDES_neutrino, ANDES}, located at $30^\circ$ South, lies outside this region but should still offer advantages compared to a similarly sized detector in the Northern Hemisphere due to its longer exposure time.

\subsubsection{Detector depths and high northern latitudes}
Although detectors located above $45^\circ$ North always see Cygnus above the horizon, those at sufficient depths can still efficiently probe the parameter space for larger $\delta$, where the short decay lengths require only a small Earth volume. Notably, Ref.~\cite{Luminous} did not calculate bounds for the SNO+ experiment~\cite{SNO+} due to its latitude at $46.5^\circ$ North. It is an otherwise promising large-volume detector, with a radius of $6$ m, slightly larger than Borexino. It could, in principle, achieve a threshold as low as $\sim 200 - 400$ keV~\cite{SNO+Threshold}, and our signal motivates pushing it toward the lower value. As we show below, despite its high latitude, its $2.07$~km depth makes it viable for the relevant range of $\delta$.

For an underground detector at depth $d_D$ below Earth's surface in the Northern Hemisphere with a latitude of $\theta_{D}$, let $\alpha_D \equiv \theta_D-45^\circ$, which can be either positive or negative. Define the maximum Earth-penetration length $\hmax$ as the distance a DM particle from Cygnus must traverse through the Earth to reach the detector when the detector is furthest from Cygnus. It is given by
\begin{align} \label{eq: hmax}
    \hmax &= \REarth \sqrt{ \sin^2\alpha_D + 2 \frac{d_D}{\REarth} \cos^2 \alpha_D } - \REarth \sin\alpha_D ~,
\end{align}
where $\REarth=6371$ km is the radius of the Earth.

Given a value of $\hmax$ and an element $A$ in the Earth, there exists a corresponding minimum mass splitting $\deltaminA$ that can be efficiently probed. This can be determined by setting the decay length $\lH=\hmax$ in \Eq{\ref{eq: decay length}} and using the minimum velocity $\vglobalmin$ needed to upscatter element $A$ from \Eq{\ref{eq: vmin global}}:
\begin{align} \label{eq: deltaminA}
\deltaminA = 87~\text{keV} \left(\frac{50~\text{GeV}}{\mu_{A}}\right) ^{1/5} \left(\frac{1000~\text{km}}{\hmax}\right)^{2/5}~.
\end{align}
Below this $\deltaminA$, the signal does not drop off exponentially, but the available Earth volume suffers a power-law suppression. Specifically, the effective volume of Earth that can produce signals, $\Veff$ in \Eq{\ref{eq:Veff estimate}}, is overestimated. We need to modify it by a suppression factor,
\begin{align}
\Veff \to \Veff \times \dfrac{\hmax}{\lH} ~,
\end{align}
where $\lH$ is a function of $\delta$ and its velocity dependence can be taken to be the $\vglobalmin$ for element $A$. In practice, at these low values of $\delta$, we only need to focus on Fe due to its large abundance. 

Table~\ref{tab: detectors} shows that, due to their lower latitudes, Borexino, JUNO, and KamLAND have large maximum Earth penetration lengths, and their $\deltaminFe$ values are well below their experimental thresholds. In contrast, the maximum Earth-penetration length for SNO+ is \numcolor{$\hmax^{\text{SNO}}=66$ km}, which is slightly less than the farthest distance a Higgsino scattering off Fe can travel, \numcolor{$\lmaxLMC{Fe} = 70$ km} (Table \ref{tab: elements}). This implies that SNO+ cannot fully take advantage of the Fe abundances, as it suffers a suppression factor of $\dfrac{h_{\text{max}}^{\text{SNO}}}{\lH}$ below \numcolor{$\deltaminFe^{\text{SNO}} = 260$ keV}. 
However, for mass splitting above this value, which is below the current Higgsino bound set by PandaX-4T (\numcolor{$\delta^{\text{SHM}}_{\text{PandaX}} \sim 260$ keV} and \numcolor{$\delta^{\text{LMC}}_{\text{PandaX}} \sim 340$ keV}), the small Earth volume available to SNO+ does not affect its signal at all. In this regime, SNO+ performs just as well as a detector of the same size at lower latitudes.

In fact, this calculation is a conservative underestimate. We have assumed the high-speed DM particles originate solely from Cygnus, whereas in reality, there is some angular spread in both the SHM~\cite{Luminous} and the LMC model~\cite{LMCHighSpeed}. Additionally, Higgsino scattering introduces a deflection angle of approximately  $\sim 10^{\circ}$~\cite{Luminous}. These angular spreads allow for contributions from below $\alpha_D = 0$, where DM particles pass through a significantly larger Earth volume.

\section{Results and Projections\label{sec:projection}}
In this section, we present the projections of the various luminous DM detection experiments and strategies described in this paper, using both the SHM and the LMC model. In addition to the Higgsino cross section of $\sxN=10^{-39}~\text{cm}^2$, we also show results for a broader range of cross sections $\sigma_{\text{DM},n}$ that apply to more general models of inelastic DM with the same mass as the Higgsino, $\mx=1.1$ TeV, and the same decay length $\lH$ [\Eq{\ref{eq: decay length}}], though the latter condition can be relaxed for the curves without Pb/U supplements. In all cases, we use Borexino's total run time of $\ttotal=3628.7~\text{days}\approx 10~\text{years}$, although the signal is unsuppressed only half that time, $\tsignal = \ttotal/2$, due to daily modulation\footnote{As explained in Sec.~\ref{sec:supplement}, the supplemental Pb/U scatters at a constant rate, so its exposure time is not subject to this factor of $2$.}. We assume the same total runtime for other upcoming experiments. All projections scale with the square root of time due to the presence of backgrounds, meaning that running the experiment for one year would yield constraints within a factor of $\approx 3$ of the limits shown here. The only exception is the background-free experiment (brown dash-dotted line in Fig.~\ref{fig:overall JUNO}), which improves linearly with time.

We first present in Fig.~\ref{fig:Borexino} the projections for the Borexino experiment. This figure highlights the improvement enabled by the LMC model compared to the SHM; for Higgsino detection, this results in an improvement from \numcolor{$\delta^{\text{SHM}}_{\text{Borexino}}\sim 330$ keV} to \numcolor{$\delta^{\text{LMC}}_{\text{Borexino}} \sim 440 $ keV}. We also show the result for the LMC model without including the terrestrial abundances of Th and U.  The inclusion of these elements is evidently not relevant for the Higgsino line itself at Borexino (and they were not included in previous work~\cite{Luminous}). However including Th and U immediately improves bounds at higher cross sections than $\sxN=10^{-39}~\text{cm}^2$, and as we pull down the constraints curves (shown in later figures) with our improvement strategies, these elements will become important for the Higgsino as well.  For example, these elements are important to include for SNO+, JUNO, or KamLAND in the LMC model.
The limits in Fig.~\ref{fig:Borexino} are attainable with Borexino's existing $\sim 10$ year data, and we advocate for this analysis, as it represents the most cost-effective advancement currently available for Higgsino dark matter.

Several features in this plot are worth noting. Due to the large background rate coming from ${}^{14}\text{C}$ $\beta$-decay (see Fig. 6 of Ref.~\cite{Luminous}), there is a reduced sensitivity below $250$ keV, with a complete cutoff at the threshold of \numcolor{$\Ethr=150$ keV}. This reduced sensitivity is apparent in the SHM curve. However, in the LMC curve, this feature is overshadowed by a sharp downward turn at \numcolor{$\delta^{\text{LMC}}_{\text{max,Fe}} = 250$ keV}, resulting from the sudden availability of Fe for scattering below this threshold. The vast abundance of Fe---four orders of magnitude higher than Sr and Ba---enables much stronger sensitivity. In the SHM curve, due to the lower maximum velocity $\vmax^{\text{SHM}}<\vmax^{\text{LMC}}$, the Fe kink occurs at \numcolor{$\delta^{\text{SHM}}_{\text{max,Fe}}=~160$~keV}. Meanwhile, the sharp decline below \numcolor{$\delta_{\text{{mantle,Ba}}}= 300$ keV} is due to the terrestrial abundance of Ba being a factor of $\sim 40$ lower in the mantle than in the crust (see Table~\ref{tab: elements}), which we model as an abrupt transition at $30$ km depth (see Appendix~\ref{sec:terrestrial elements}). These features are consistent across all subsequent figures.

In Figs.~\ref{fig:overall SNO}--\ref{fig:overall JUNO}, we present the projected limits of the SNO+, KamLAND, and JUNO experiments, respectively, with the SHM and the LMC model shown in different panels. The Higgsino sensitivities of the standard luminous detection scheme (shown in green) of SNO+, KamLAND, and JUNO are \numcolor{$\delta^{\text{LMC}}_{\text{SNO}}\sim 460$ keV}, \numcolor{$\delta^{\text{LMC}}_{\text{KamLAND}}\sim 420$ keV}, and \numcolor{$\delta^{\text{LMC}}_{\text{JUNO}} \sim 540$ keV}, respectively; they are also shown in Fig.~\ref{fig:graphic} in green and red blocks. These curves improve over Borexino by the square root of their detector volumes, as we assume the background per scintillator mass is equal across experiments\footnote{This leads to a somewhat different projection from Ref.~\cite{Luminous}, where it was assumed that the total number of background events would be the same in JUNO as it was in Borexino, rather than the number per ton. Also, we used the published background of KamLAND~\cite{Kamland}; this results in the KamLAND projection being somewhat weaker than that of SNO+ despite being similar size, but that is solely because we made the optimistic asumption that SNO+ can reach the excellent background level of Borexino.}.

 We also show the enhanced sensitivity SNO+, KamLAND, and JUNO would achieve if a large volume of supplemental Pb (orange lines) or U (purplish-blue lines) is installed near the detector, as described in Sec.~\ref{sec:supplement}. Since this method may be limited by the available space near the detector (and potentially by the cost of excavating additional volume), we report the results based on the volume of the supplemental Pb/U, considering the benchmark volumes of $(5~\text{m})^3$ and $(20~\text{m})^3$. Pb has a cutoff point of \numcolor{$\delta^{\text{LMC}}_{\text{max,Pb}} = 830$ keV}, beyond which the supplemental Pb no longer contributes. With $(20~\text{m})^3$ of supplemental Pb, SNO+ and JUNO have Higgsino sensitivities near this limit, with $\delta^{\text{LMC}}_{\text{SNO,Pb}}=730$ keV and $\delta^{\text{LMC}}_{\text{JUNO,Pb}}=750$ keV, while KamLAND reaches $\delta^{\text{LMC}}_{\text{SNO,Pb}}=650$ keV. With $(20~\text{m})^3$ of U, JUNO's Higgsino bound can reach \numcolor{$\delta^{\text{LMC}}_{\text{JUNO,U}} \sim 850$ keV}, while SNO+ and KamLAND follow closely at \numcolor{$\delta^{\text{LMC}}_{\text{SNO,U}} \sim 810$~keV} and \numcolor{$\delta^{\text{LMC}}_{\text{KamLAND,U}} \sim 750$ keV}, respectively. Encouragingly, even just $(5~\text{m})^3$ of U still yields significant gains. These limits are shown in Fig.~\ref{fig:graphic} in purple and brown, labeled ``+Pb" and ``+U."

We also include in Fig.~\ref{fig:overall JUNO} projections of a background-free $\gamma$-ray detector (shown by the brown dash-dotted line) with the same $17.5$-meter radius as JUNO, as discussed in Section~\ref{sec:BGF}. Such a detector can probe splittings up to \numcolor{$\delta^{\text{LMC}}_{\text{BGF}}\sim 800 $ keV} for the Higgsino. We emphasize that this is a preliminary idea, and further research is needed to determine the feasibility of such a background-free detector. While complete elimination of background may not be achievable in practice, we present this background-free curve as an ultimate limit. Additionally, supplemental U could be installed around the background-free detector. In this most optimistic scenario, $(20~\text{m})^3$ of U (not shown) is sufficient to fully saturate the \numcolor{$\delta^{\text{LMC}}_{\text{max,U}}\sim 940$ keV} limit. 

\FloatBarrier
\section{Conclusion}
In this work, we studied the feasibility of terrestrial detection for one of the few theoretically well-motivated WIMPs that have still survived: the pseudo-Dirac Higgsino. Its splitting has been constrained to \numcolor{$\delta \gtrsim 260$~keV} from traditional direct detection in existing literature, assuming the SHM for dark matter velocity. In this work, we showed that incorporating the additional fast component of dark matter, due to the merger of the LMC with the MW, improves these constraints to larger splitting, \numcolor{$\delta \gtrsim 340~\textrm{keV}$}.  We also noted that a luminous detection strategy, where Higgsinos scatter off the Earth and subsequently decay in large-volume neutrino detectors producing a monoenergetic photon, shows great promise for pushing to much higher $\delta$.
Such a detection strategy also benefits from the faster dark matter in the LMC halo model, as well as the presence of heavy elements such as U and Th in the Earth.
A reanalysis of the 10-year Borexino dataset searching for this luminous signal could probe Higgsinos with splittings up to \numcolor{$\sim 440$~keV}.  Larger detectors, such as SNO+, KamLAND, and JUNO, show exciting promise due to their larger fiducial volumes. Additionally, placing large amounts of Pb or U next to these detectors could further improve sensitivity, reaching splittings of up to \numcolor{$\sim 900$~keV}.

It is important to note that probing splittings in the hundreds of keV range corresponds to gauginos with masses above the PeV scale. This mass range is beyond the reach of even future colliders or precision EDM experiments, and thus, it is not expected to be probed by other means in the near future. The methods introduced in this paper provide a complementary probe of smaller mass splittings, with splittings much larger than 10 MeV being expected to be probed by Advanced ACME and direct detection via Higgs scattering (see Fig.~\ref{fig:graphic}). Even future projections are going to leave a tantalizing unconstrained window between $1$~MeV and $30$~MeV. Finding novel ideas to probe this window for one of the last surviving WIMP candidates is an open problem.  

We emphasize that the $1.1$~TeV Higgsino is not the sole beneficiary of the ideas introduced in this work, namely, incorporation of the high-speed LMC dark matter population, consideration of heavy elements like Th and U in the Earth's crust, and supplementation with large blocks of heavy metal. Any inelastic dark matter candidate with mass splitting and subsequent decay of the excited state into Standard Model particles can be searched for with this technique, as long as the decay length is macroscopic and smaller than the size of the Earth. Models such as Higgsino dark matter with masses different from $1.1$~TeV, which might arise from nonstandard cosmologies or other electroweak $n$-plets with nonzero hypercharge, are excellent examples. Another candidate is the magnetic inelastic dark matter scenario. We will discuss limits and projections for these models in forthcoming work. 

\FloatBarrier
\textit{\\Note added.} After this paper appeared on arXiv, Ref.~\cite{AlbertaPaper} was posted, which partially overlaps with our work in calculating the effect of the LMC on collisional direct detection experiments. However, version 1 reported a sensitivity of only $\delta \sim 240$ keV for the proposed DARWIN experiment, which is lower than our estimate of $\delta \sim 400$ keV for PANDAX-4T. This discrepancy arises from their assumption of a nuclear recoil detection efficiency cutoff at only 21~keV, based on an older paper~\cite{OutdatedSensitivity}, whereas we used the reported efficiency by PANDAX-4T~\cite{PandaX4T2021}, which has a cutoff above 120~keV. The authors of Ref.~\cite{AlbertaPaper} have confirmed in private communication that using the updated efficiency led them to a sensitivity estimate consistent with ours.

\acknowledgments
We thank Spencer Axani, Michael Fedderke, Patrick Fox, Christopher Grant, Giorgio Gratta, Kuunal Mahtani, Jeff Tseng, Liangjian Wen, Lindley Winslow, and Josephine Wong for helpful discussions. We also thank the anonymous reviewer for helpful comments.

The authors acknowledge support by NSF Grants No.~PHY-2310429, Simons Investigator Award No.~824870, DOE HEP QuantISED award No.~100495, the Gordon and Betty Moore Foundation Grant No.~GBMF7946, and the U.S. Department of Energy (DOE), Office of Science, National Quantum Information Science Research Centers, Superconducting Quantum Materials and Systems Center (SQMS) under contract No.~DEAC02-07CH11359. S.W.~was supported in part by the J.J., L.P., and A.J. Smortchevsky Fellowship.

\renewcommand{\appendixname}{APPENDIX}
\appendix
\section{TERRESTRIAL ELEMENT ABUNDANCES\label{sec:terrestrial elements}}
Our sensitivity is set by the number densities $n_A$ of the elements in the Earth, but they are often reported in mass abundances $a_A$ (i.e., mass fraction). The two quantities are related by
\begin{align}
    n_A = a_A \times \frac{N_0}{M_A} \rho ~,
\end{align}
where $N_0$ is Avogadro's number, $M_A$ is the molar mass, and $\rho$ is the density of the Earth. However, the Earth's density and the elemental abundances vary with depth. For this calculation, we need only consider the crust and mantle, which have average densities $\rho_{\text{crust}}=2830 ~\text{kg}/\text{m}^3$~\cite{crust_density} and $\rho_{\text{mantle}}=4500 ~\text{kg}/\text{m}^3$~\cite{mantleDensity}. We will also take into account the elemental abundances in conjunction with the characteristic decay lengths $\lmaxA$ [\Eq{\ref{eq:lmaxA}}].

The first $\sim 30$ km below Earth's surface constitutes the Earth's crust, which can be divided into the upper, middle, and lower crust, each with different elemental abundances. Instead of treating all three layers separately, we note that for the four heaviest elements listed in Table \ref{tab: elements}, each has significantly lower abundances than the next heaviest element on our list, making their contributions important only above the $\deltamaxA$ of the preceding element. This simplifies our calculation, as we can safely assume most of these elements have constant abundances. In effect, since the upper crust is estimated to be $\sim 15$ km thick~\cite{Composition_Crust}, and Ba has a characteristic decay length (\numcolor{$\lmaxLMC{Ba} = 5.8$ km}) well within the upper crust, we can use the upper-crust abundances of Pb, Th, and U, even though these elements are less abundant in the middle and lower crust~\cite{Composition_Crust}.

Sr and Ba have appreciable abundances in both the mantle and the crust, but their number densities in the mantle are about an order of magnitude lower than their values in the crust. To account for this, we compute the $\delta$ at which the corresponding decay length reaches $30$ km depth, using the minimum velocity required to scatter the given element [\Eq{\ref{eq: vmin global}}]. This calculation yields
\begin{align}
    \delta_{\text{mantle,A}} = 300~\text{keV} \left(\frac{117~\text{GeV}}{\mu_A}\right)^{1/5} ~.
\end{align}
For instance, for \numcolor{$\delta<\delta_{\text{mantle,Ba}}=300$ keV}, we use Ba’s mantle number density; otherwise, we use its crust density. We do not need to repeat this calculation for Ca and Fe, since \numcolor{$\delta_{\text{mantle,Ca}}=380~\text{keV}$} and \numcolor{$\delta_{\text{mantle,Fe}}=360~\text{keV}$} are well above \numcolor{$\deltamaxLMC{Ca} = 180~\text{keV}$} and \numcolor{$\deltamaxLMC{Fe} = 250~\text{keV}$}; see Table~\ref{tab: elements}. Thus, only their mantle abundances are relevant.

\bibliography{refs}

@article{Hall:1997ah,
    author = "Hall, Lawrence J. and Moroi, Takeo and Murayama, Hitoshi",
    title = "{Sneutrino cold dark matter with lepton number violation}",
    eprint = "hep-ph/9712515",
    archivePrefix = "arXiv",
    reportNumber = "LBNL-41199, LBL-41199, UCB-PTH-97-69",
    doi = "10.1016/S0370-2693(98)00196-8",
    journal = "Phys. Lett. B",
    volume = "424",
    pages = "305--312",
    year = "1998"
}

@article{Tucker-Smith:2001myb,
    author = "Tucker-Smith, David and Weiner, Neal",
    title = "{Inelastic dark matter}",
    eprint = "hep-ph/0101138",
    archivePrefix = "arXiv",
    reportNumber = "UCB-PTH-00-43, LBNL-47234, UW-PT-00-17",
    doi = "10.1103/PhysRevD.64.043502",
    journal = "Phys. Rev. D",
    volume = "64",
    pages = "043502",
    year = "2001"
}

@article{Barn,
    author = "Digman, Matthew C. and Cappiello, Christopher V. and Beacom, John F. and Hirata, Christopher M. and Peter, Annika H. G.",
    title = "{Not as big as a barn: Upper bounds on dark matter-nucleus cross sections}",
    eprint = "1907.10618",
    archivePrefix = "arXiv",
    primaryClass = "hep-ph",
    doi = "10.1103/PhysRevD.100.063013",
    journal = "Phys. Rev. D",
    volume = "100",
    number = "6",
    pages = "063013",
    year = "2019",
    note = "[Erratum: Phys.Rev.D 106, 089902 (2022)]"
}

@article{Frontier,
    author = "Song, Ningqiang and Nagorny, Serge and Vincent, Aaron C.",
    title = "{Pushing the frontier of WIMPy inelastic dark matter: Journey to the end of the periodic table}",
    eprint = "2104.09517",
    archivePrefix = "arXiv",
    primaryClass = "hep-ph",
    doi = "10.1103/PhysRevD.104.103032",
    journal = "Phys. Rev. D",
    volume = "104",
    number = "10",
    pages = "103032",
    year = "2021"
}

@article{LMC,
    author = "Smith-Orlik, Adam and others",
    title = "{The impact of the Large Magellanic Cloud on dark matter direct detection signals}",
    eprint = "2302.04281",
    archivePrefix = "arXiv",
    primaryClass = "astro-ph.GA",
    doi = "10.1088/1475-7516/2023/10/070",
    journal = "JCAP",
    volume = "10",
    pages = "070",
    year = "2023"
}

@article{Luminous,
    author = "Eby, Joshua and Fox, Patrick J. and Harnik, Roni and Kribs, Graham D.",
    title = "{Luminous signals of inelastic dark matter in large detectors}",
    eprint = "1904.09994",
    archivePrefix = "arXiv",
    primaryClass = "hep-ph",
    reportNumber = "FERMILAB-PUB-19-147-T",
    doi = "10.1007/JHEP09(2019)115",
    journal = "JHEP",
    volume = "09",
    pages = "115",
    year = "2019"
}

@article{Pospelov:2013nea,
    author = "Pospelov, Maxim and Weiner, Neal and Yavin, Itay",
    title = "{Dark matter detection in two easy steps}",
    eprint = "1312.1363",
    archivePrefix = "arXiv",
    primaryClass = "hep-ph",
    doi = "10.1103/PhysRevD.89.055008",
    journal = "Phys. Rev. D",
    volume = "89",
    number = "5",
    pages = "055008",
    year = "2014"
}

@article{Borexino_Phase_I,
    author = "Bellini, G. and others",
    collaboration = "Borexino",
    title = "{Final results of Borexino Phase-I on low energy solar neutrino spectroscopy}",
    eprint = "1308.0443",
    archivePrefix = "arXiv",
    primaryClass = "hep-ex",
    doi = "10.1103/PhysRevD.89.112007",
    journal = "Phys. Rev. D",
    volume = "89",
    number = "11",
    pages = "112007",
    year = "2014"
}

@article{Borexino_Electron_Decay,
    author = "Agostini, M. and others",
    collaboration = "Borexino",
    title = "{A test of electric charge conservation with Borexino}",
    eprint = "1509.01223",
    archivePrefix = "arXiv",
    primaryClass = "hep-ex",
    doi = "10.1103/PhysRevLett.115.231802",
    journal = "Phys. Rev. Lett.",
    volume = "115",
    pages = "231802",
    year = "2015"
}

@article{Borexino_Latest,
    author = "Basilico, D. and others",
    collaboration = "BOREXINO",
    title = "{Final results of Borexino on CNO solar neutrinos}",
    eprint = "2307.14636",
    archivePrefix = "arXiv",
    primaryClass = "hep-ex",
    doi = "10.1103/PhysRevD.108.102005",
    journal = "Phys. Rev. D",
    volume = "108",
    number = "10",
    pages = "102005",
    year = "2023"
}

@article{LuminousDM,
    author = "Feldstein, Brian and Graham, Peter W. and Rajendran, Surjeet",
    title = "{Luminous dark matter}",
    eprint = "1008.1988",
    archivePrefix = "arXiv",
    primaryClass = "hep-ph",
    reportNumber = "MIT-CTP-4172",
    doi = "10.1103/PhysRevD.82.075019",
    journal = "Phys. Rev. D",
    volume = "82",
    pages = "075019",
    year = "2010"
}

@article{LMCHighSpeed,
    author = "Besla, Gurtina and Peter, Annika and Garavito-Camargo, Nicolas",
    title = "{The highest-speed local dark matter particles come from the Large Magellanic Cloud}",
    eprint = "1909.04140",
    archivePrefix = "arXiv",
    primaryClass = "astro-ph.GA",
    doi = "10.1088/1475-7516/2019/11/013",
    journal = "JCAP",
    volume = "11",
    pages = "013",
    year = "2019"
}

@article{Original_SHM,
  title = {Detecting cold dark-matter candidates},
  author = {Drukier, Andrzej K. and Freese, Katherine and Spergel, David N.},
  journal = {Phys. Rev. D},
  volume = {33},
  issue = {12},
  pages = {3495--3508},
  numpages = {0},
  year = {1986},
  month = {Jun},
  publisher = {American Physical Society},
  doi = {10.1103/PhysRevD.33.3495},
  url = {https://link.aps.org/doi/10.1103/PhysRevD.33.3495}
}

@article{SHM++,
    author = "Evans, N. Wyn and O'Hare, Ciaran A. J. and McCabe, Christopher",
    title = "{Refinement of the standard halo model for dark matter searches in light of the Gaia Sausage}",
    eprint = "1810.11468",
    archivePrefix = "arXiv",
    primaryClass = "astro-ph.GA",
    doi = "10.1103/PhysRevD.99.023012",
    journal = "Phys. Rev. D",
    volume = "99",
    number = "2",
    pages = "023012",
    year = "2019"
}

@inproceedings{Lisanti_TASI,
    author = "Lisanti, Mariangela",
    title = "{Lectures on Dark Matter Physics}",
    eprint = "1603.03797",
    archivePrefix = "arXiv",
    primaryClass = "hep-ph",
    doi = "10.1142/9789813149441_0007",
    year = "2017"
}

@article{SMH_Deviation_1,
    author = "Evans, N. Wyn and An, Jin H.",
    title = "{Distribution function of the dark matter}",
    eprint = "astro-ph/0511687",
    archivePrefix = "arXiv",
    doi = "10.1103/PhysRevD.73.023524",
    journal = "Phys. Rev. D",
    volume = "73",
    pages = "023524",
    year = "2006"
}

@article{SMH_Deviation_2,
    author = "Vogelsberger, Mark and Helmi, A. and Springel, Volker and White, Simon D. M. and Wang, Jie and Frenk, Carlos S. and Jenkins, Adrian and Ludlow, A. D. and Navarro, Julio F.",
    title = "{Phase-space structure in the local dark matter distribution and its signature in direct detection experiments}",
    eprint = "0812.0362",
    archivePrefix = "arXiv",
    primaryClass = "astro-ph",
    doi = "10.1111/j.1365-2966.2009.14630.x",
    journal = "Mon. Not. Roy. Astron. Soc.",
    volume = "395",
    pages = "797--811",
    year = "2009"
}

@article{SMH_Deviation_3,
    author = "Zemp, Marcel and Diemand, Jurg and Kuhlen, Michael and Madau, Piero and Moore, Ben and Potter, Doug and Stadel, Joachim and Widrow, Lawrence",
    title = "{The Graininess of Dark Matter Haloes}",
    eprint = "0812.2033",
    archivePrefix = "arXiv",
    primaryClass = "astro-ph",
    doi = "10.1111/j.1365-2966.2008.14361.x",
    journal = "Mon. Not. Roy. Astron. Soc.",
    volume = "394",
    pages = "641--659",
    year = "2009"
}

@article{SMH_Deviation_4,
    author = "Kuhlen, Michael and Weiner, Neal and Diemand, Jurg and Madau, Piero and Moore, Ben and Potter, Doug and Stadel, Joachim and Zemp, Marcel",
    title = "{Dark Matter Direct Detection with Non-Maxwellian Velocity Structure}",
    eprint = "0912.2358",
    archivePrefix = "arXiv",
    primaryClass = "astro-ph.GA",
    doi = "10.1088/1475-7516/2010/02/030",
    journal = "JCAP",
    volume = "02",
    pages = "030",
    year = "2010"
}

@article{SMH_Deviation_5,
    author = "Mao, Yao-Yuan and Strigari, Louis E. and Wechsler, Risa H.",
    title = "{Connecting Direct Dark Matter Detection Experiments to Cosmologically Motivated Halo Models}",
    eprint = "1304.6401",
    archivePrefix = "arXiv",
    primaryClass = "astro-ph.CO",
    doi = "10.1103/PhysRevD.89.063513",
    journal = "Phys. Rev. D",
    volume = "89",
    number = "6",
    pages = "063513",
    year = "2014"
}

@article{LMC_Similar,
   title={Effects on the local dark matter distribution due to the large magellanic cloud},
   volume={513},
   ISSN={1745-3933},
   url={http://dx.doi.org/10.1093/mnrasl/slac031},
   DOI={10.1093/mnrasl/slac031},
   number={1},
   journal={Monthly Notices of the Royal Astronomical Society: Letters},
   publisher={Oxford University Press (OUP)},
   author={Donaldson, Katelin and Petersen, Michael S and Peñarrubia, Jorge},
   year={2022},
   month=mar, pages={46–51} }

@article{Directional_DM,
    author = "Grothaus, Philipp and Fairbairn, Malcolm and Monroe, Jocelyn",
    title = "{Directional Dark Matter Detection Beyond the Neutrino Bound}",
    eprint = "1406.5047",
    archivePrefix = "arXiv",
    primaryClass = "hep-ph",
    reportNumber = "KCL-PH-TH-2014-28, LCTS-2014-27",
    doi = "10.1103/PhysRevD.90.055018",
    journal = "Phys. Rev. D",
    volume = "90",
    number = "5",
    pages = "055018",
    year = "2014"
}

@article{Auriga,
    author = "Grand, Robert J. J. and G\'omez, Facundo A. and Marinacci, Federico and Pakmor, Ruediger and Springel, Volker and Campbell, David J. R. and Frenk, Carlos S. and Jenkins, Adrian and White, Simon D. M.",
    title = "{The Auriga Project: the properties and formation mechanisms of disc galaxies across cosmic time}",
    eprint = "1610.01159",
    archivePrefix = "arXiv",
    primaryClass = "astro-ph.GA",
    doi = "10.1093/mnras/stx071",
    journal = "Mon. Not. Roy. Astron. Soc.",
    volume = "467",
    number = "1",
    pages = "179--207",
    year = "2017"
}

@article{FirstInfall,
    author = "Besla, Gurtina and Kallivayalil, Nitya and Hernquist, Lars and Robertson, Brant and Cox, T. J. and van der Marel, Roeland P. and Alcock, Charles",
    title = "{Are the Magellanic Clouds on their First Passage about the Milky Way?}",
    eprint = "astro-ph/0703196",
    archivePrefix = "arXiv",
    doi = "10.1086/521385",
    journal = "Astrophys. J.",
    volume = "668",
    pages = "949--967",
    year = "2007"
}

@article{MWMass,
    author = "Klypin, Anatoly and Zhao, HongSheng and Somerville, Rachel S.",
    title = "{Lambda CDM-based models for the Milky Way and M31 I: Dynamical models}",
    eprint = "astro-ph/0110390",
    archivePrefix = "arXiv",
    doi = "10.1086/340656",
    journal = "Astrophys. J.",
    volume = "573",
    pages = "597--613",
    year = "2002"
}

@article{DarkMatterDensity,
    author = "Read, J. I.",
    title = "{The Local Dark Matter Density}",
    eprint = "1404.1938",
    archivePrefix = "arXiv",
    primaryClass = "astro-ph.GA",
    reportNumber = "JPHYSG-100038.R1",
    doi = "10.1088/0954-3899/41/6/063101",
    journal = "J. Phys. G",
    volume = "41",
    pages = "063101",
    year = "2014"
}

@article{LMC_Velocity,
   title={RAM PRESSURE STRIPPING OF THE LARGE MAGELLANIC CLOUD’S DISK AS A PROBE OF THE MILKY WAY’S CIRCUMGALACTIC MEDIUM},
   volume={815},
   ISSN={1538-4357},
   url={http://dx.doi.org/10.1088/0004-637X/815/1/77},
   DOI={10.1088/0004-637x/815/1/77},
   number={1},
   journal={The Astrophysical Journal},
   publisher={American Astronomical Society},
   author={Salem, Munier and Besla, Gurtina and Bryan, Greg and Putman, Mary and van der Marel, Roeland P. and Tonnesen, Stephanie},
   year={2015},
   month=dec, pages={77} }

@article{crust_density,
author = {Christensen, Nikolas I. and Mooney, Walter D.},
title = {Seismic velocity structure and composition of the continental crust: A global view},
journal = {Journal of Geophysical Research: Solid Earth},
volume = {100},
number = {B6},
pages = {9761-9788},
doi = {https://doi.org/10.1029/95JB00259},
url = {https://agupubs.onlinelibrary.wiley.com/doi/abs/10.1029/95JB00259},
year = {1995}
}

@incollection{Composition_Crust,
title = {4.1 - Composition of the Continental Crust},
editor = {Heinrich D. Holland and Karl K. Turekian},
booktitle = {Treatise on Geochemistry (Second Edition)},
publisher = {Elsevier},
edition = {Second Edition},
address = {Oxford},
pages = {1-51},
year = {2014},
isbn = {978-0-08-098300-4},
doi = {https://doi.org/10.1016/B978-0-08-095975-7.00301-6},
url = {https://www.sciencedirect.com/science/article/pii/B9780080959757003016},
author = {R.L. Rudnick and S. Gao}
}

@incollection{Mantle_Composition,
title = {3.1 - Cosmochemical Estimates of Mantle Composition},
editor = {Heinrich D. Holland and Karl K. Turekian},
booktitle = {Treatise on Geochemistry (Second Edition)},
publisher = {Elsevier},
edition = {Second Edition},
address = {Oxford},
pages = {1-39},
year = {2014},
isbn = {978-0-08-098300-4},
doi = {https://doi.org/10.1016/B978-0-08-095975-7.00201-1},
author = {H. Palme and H.St.C. O'Neill}
}

@misc{ANDES,
      title={The ANDES Deep Underground Laboratory}, 
      author={X. Bertou},
      year={2013},
      eprint={1308.0059},
      archivePrefix={arXiv},
      primaryClass={astro-ph.IM},
      url={https://arxiv.org/abs/1308.0059}, 
}

@article{ANDES_neutrino,
    author = "Machado, P. A. N. and Muhlbeier, T. and Nunokawa, H. and Zukanovich Funchal, R.",
    title = "{Potential of a Neutrino Detector in the ANDES Underground Laboratory for Geophysics and Astrophysics of Neutrinos}",
    eprint = "1207.5454",
    archivePrefix = "arXiv",
    primaryClass = "hep-ph",
    doi = "10.1103/PhysRevD.86.125001",
    journal = "Phys. Rev. D",
    volume = "86",
    pages = "125001",
    year = "2012"
}

@article{SNO+,
    author = "Albanese, V. and others",
    collaboration = "SNO+",
    title = "{The SNO+ experiment}",
    eprint = "2104.11687",
    archivePrefix = "arXiv",
    primaryClass = "physics.ins-det",
    doi = "10.1088/1748-0221/16/08/P08059",
    journal = "JINST",
    volume = "16",
    number = "08",
    pages = "P08059",
    year = "2021"
}

@inproceedings{SNO+Depth,
    author = "Simms, Jasmine",
    title = "{Muon Track Reconstruction in the Scintillator Phase of SNO+}",
    booktitle = "{Prospects in Neutrinos Physics}",
    eprint = "2404.03680",
    archivePrefix = "arXiv",
    primaryClass = "physics.ins-det",
    month = "3",
    year = "2024"
}

@article{SNO+Threshold,
    author = "Andringa, S. and others",
    collaboration = "SNO+",
    title = "{Current Status and Future Prospects of the SNO+ Experiment}",
    eprint = "1508.05759",
    archivePrefix = "arXiv",
    primaryClass = "physics.ins-det",
    doi = "10.1155/2016/6194250",
    journal = "Adv. High Energy Phys.",
    volume = "2016",
    pages = "6194250",
    year = "2016"
}

@article{JUNO_Depth,
author ={{JUNO Collaboration}},
title = {JUNO physics and detector},
journal = {Progress in Particle and Nuclear Physics},
volume = {123},
pages = {103927},
year = {2022},
issn = {0146-6410},
doi = {https://doi.org/10.1016/j.ppnp.2021.103927},
url = {https://www.sciencedirect.com/science/article/pii/S0146641021000880}
}

@misc{JUNO_Conceptual,
    author = "Djurcic, Zelimir and others",
    collaboration = "JUNO",
    title = "{JUNO Conceptual Design Report}",
    eprint = "1508.07166",
    archivePrefix = "arXiv",
    primaryClass = "physics.ins-det",
    month = "8",
    year = "2015"
}

@article{RadiativeDecay,
title = {Radiative neutralino decay},
journal = {Nuclear Physics B},
volume = {323},
number = {2},
pages = {267-310},
year = {1989},
issn = {0550-3213},
doi = {https://doi.org/10.1016/0550-3213(89)90143-0},
url = {https://www.sciencedirect.com/science/article/pii/0550321389901430},
author = {Howard E. Haber and Daniel Wyler}
}

@misc{StatsBook,
author = "Leo, W. R.",
title = "Techniques for Nuclear and Particle Physics Experiments",
year="1987"
}

@article{acme2018improved,
	title = {Improved limit on the electric dipole moment of the electron},
	volume = {562},
	issn = {1476-4687},
	url = {https://doi.org/10.1038/s41586-018-0599-8},
	doi = {10.1038/s41586-018-0599-8},
	number = {7727},
	journal = {Nature},
	author = {Andreev, V. and Ang, D. G. and DeMille, D. and Doyle, J. M. and Gabrielse, G. and Haefner, J. and Hutzler, N. R. and Lasner, Z. and Meisenhelder, C. and O’Leary, B. R. and Panda, C. D. and West, A. D. and West, E. P. and Wu {(ACME Collaboration)}, X.},
	month = oct,
	year = {2018},
	pages = {355--360},
}

@article{PandaXII2022,
    author = "Yuan, Ying and others",
    collaboration = "PandaX",
    title = "{A search for two-component Majorana dark matter in a simplified model using the full exposure data of PandaX-II experiment}",
    eprint = "2205.08066",
    archivePrefix = "arXiv",
    primaryClass = "hep-ex",
    doi = "10.1016/j.physletb.2022.137254",
    journal = "Phys. Lett. B",
    volume = "832",
    pages = "137254",
    year = "2022"
}

@article{PandaX4T2021,
  title = {Dark Matter Search Results from the PandaX-4T Commissioning Run},
  author = {Meng \textit{et al.}, Yue},
  collaboration = {PandaX-4T Collaboration},
  journal = {Phys. Rev. Lett.},
  volume = {127},
  issue = {26},
  pages = {261802},
  numpages = {8},
  year = {2021},
  month = {Dec},
  publisher = {American Physical Society},
  doi = {10.1103/PhysRevLett.127.261802},
  url = {https://link.aps.org/doi/10.1103/PhysRevLett.127.261802}
}

@article{PICO2023,
    author = "Adams, E. and others",
    collaboration = "PICO Collaboration",
    title = "{Search for inelastic dark matter-nucleus scattering with the PICO-60 CF3I and C3F8 bubble chambers}",
    eprint = "2301.08993",
    archivePrefix = "arXiv",
    primaryClass = "astro-ph.CO",
    reportNumber = "FERMILAB-PUB-23-059-PPD",
    doi = "10.1103/PhysRevD.108.062003",
    journal = "Phys. Rev. D",
    volume = "108",
    number = "6",
    pages = "062003",
    year = "2023"
}

@article{Co:2021ion,
    author = "Co, Raymond T. and Sheff, Benjamin and Wells, James D.",
    title = "{Race to find split Higgsino dark matter}",
    eprint = "2105.12142",
    archivePrefix = "arXiv",
    primaryClass = "hep-ph",
    reportNumber = "LCTP-21-11, UMN-TH-4016/21, FTPI-MINN-21/09",
    doi = "10.1103/PhysRevD.105.035012",
    journal = "Phys. Rev. D",
    volume = "105",
    number = "3",
    pages = "035012",
    year = "2022"
}

@article{LastWIMP,
    author = "Krall, Rebecca and Reece, Matthew",
    title = "{Last Electroweak WIMP Standing: Pseudo-Dirac Higgsino Status and Compact Stars as Future Probes}",
    eprint = "1705.04843",
    archivePrefix = "arXiv",
    primaryClass = "hep-ph",
    doi = "10.1088/1674-1137/42/4/043105",
    journal = "Chin. Phys. C",
    volume = "42",
    number = "4",
    pages = "043105",
    year = "2018"
}

@article{isomer1,
    author = "Pospelov, Maxim and Rajendran, Surjeet and Ramani, Harikrishnan",
    title = "{Metastable Nuclear Isomers as Dark Matter Accelerators}",
    eprint = "1907.00011",
    archivePrefix = "arXiv",
    primaryClass = "hep-ph",
    doi = "10.1103/PhysRevD.101.055001",
    journal = "Phys. Rev. D",
    volume = "101",
    number = "5",
    pages = "055001",
    year = "2020"
}

@article{isomer4,
    author = "Alves, D. S. M. and Elliott, S. R. and Massarczyk, R. and Meijer, S. J. and Ramani, H.",
    title = "{Dark Matter Constraints from Isomeric Hf178m}",
    eprint = "2306.04442",
    archivePrefix = "arXiv",
    primaryClass = "nucl-ex",
    doi = "10.1103/PhysRevLett.131.141801",
    journal = "Phys. Rev. Lett.",
    volume = "131",
    number = "14",
    pages = "141801",
    year = "2023"
}

@article{isomer2,
    author = {Lehnert, Bj\"orn and Ramani, Harikrishnan and Hult, Mikael and Lutter, Guillaume and Pospelov, Maxim and Rajendran, Surjeet and Zuber, Kai},
    title = "{Search for Dark Matter Induced Deexcitation of $^{180}$Ta$\rm ^m$}",
    eprint = "1911.07865",
    archivePrefix = "arXiv",
    primaryClass = "astro-ph.CO",
    doi = "10.1103/PhysRevLett.124.181802",
    journal = "Phys. Rev. Lett.",
    volume = "124",
    number = "18",
    pages = "181802",
    year = "2020"
}

@article{isomer3,
    author = "Arnquist, I. J. and others",
    collaboration = "Majorana",
    title = "{Constraints on the Decay of Ta180m}",
    eprint = "2306.01965",
    archivePrefix = "arXiv",
    primaryClass = "nucl-ex",
    doi = "10.1103/PhysRevLett.131.152501",
    journal = "Phys. Rev. Lett.",
    volume = "131",
    number = "15",
    pages = "152501",
    year = "2023"
}

@article{Rodd:2024qsi,
    author = "Rodd, Nicholas L. and Safdi, Benjamin R. and Xu, Weishuang Linda",
    title = "{CTA and SWGO can discover Higgsino dark matter annihilation}",
    eprint = "2405.13104",
    archivePrefix = "arXiv",
    primaryClass = "hep-ph",
    doi = "10.1103/PhysRevD.110.043003",
    journal = "Phys. Rev. D",
    volume = "110",
    number = "4",
    pages = "043003",
    year = "2024"
}

@article{CTAConsortium:2010umy,
    author = "Actis, M. and others",
    collaboration = "CTA Consortium",
    title = "{Design concepts for the Cherenkov Telescope Array CTA: An advanced facility for ground-based high-energy gamma-ray astronomy}",
    eprint = "1008.3703",
    archivePrefix = "arXiv",
    primaryClass = "astro-ph.IM",
    doi = "10.1007/s10686-011-9247-0",
    journal = "Exper. Astron.",
    volume = "32",
    pages = "193--316",
    year = "2011"
}

@article{Rinchiuso:2020skh,
    author = "Rinchiuso, Lucia and Macias, Oscar and Moulin, Emmanuel and Rodd, Nicholas L. and Slatyer, Tracy R.",
    title = "{Prospects for detecting heavy WIMP dark matter with the Cherenkov Telescope Array: The Wino and Higgsino}",
    eprint = "2008.00692",
    archivePrefix = "arXiv",
    primaryClass = "astro-ph.HE",
    reportNumber = "MIT-CTP 5120, IRFU-20-13",
    doi = "10.1103/PhysRevD.103.023011",
    journal = "Phys. Rev. D",
    volume = "103",
    number = "2",
    pages = "023011",
    year = "2021"
}

@article{aaboud2018search,
  title={Search for electroweak production of supersymmetric states in scenarios with compressed mass spectra at s= 13 TeV with the ATLAS detector},
  author={Aaboud, Morad and Aad, Georges and Abbott, Brad and Abdinov, Ovsat and Abeloos, Baptiste and Abidi, Syed Haider and AbouZeid, OS and Abraham, Nadine L and Abramowicz, Halina and Abreu, Henso and others},
  journal={Physical Review D},
  volume={97},
  number={5},
  pages={052010},
  year={2018},
  publisher={APS}
}

@book{MantleDensity,
  title     = "Introduction to Oceanography",
  author    = "Webb, Paul",
  year      = 2023,
  publisher = "Pressbooks",
  url       = "https://rwu.pressbooks.pub/webboceanography/"
}

@article{InelasticFrontier,
    author = "Bramante, Joseph and Fox, Patrick J. and Kribs, Graham D. and Martin, Adam",
    title = "{Inelastic frontier: Discovering dark matter at high recoil energy}",
    eprint = "1608.02662",
    archivePrefix = "arXiv",
    primaryClass = "hep-ph",
    reportNumber = "FERMILAB-PUB-16-301-T",
    doi = "10.1103/PhysRevD.94.115026",
    journal = "Phys. Rev. D",
    volume = "94",
    number = "11",
    pages = "115026",
    year = "2016"
}

@article{Astrogam,
   title={Gamma-ray astrophysics in the MeV range: The ASTROGAM concept and beyond},
   volume={51},
   ISSN={1572-9508},
   url={http://dx.doi.org/10.1007/s10686-021-09706-y},
   DOI={10.1007/s10686-021-09706-y},
   number={3},
   journal={Experimental Astronomy},
   publisher={Springer Science and Business Media LLC},
   author={De Angelis \textit{et al.}, Alessandro},
   year={2021},
   month=jun, pages={1225–1254} }

@misc{CYGNUS,
    author = "Mazzitelli, G. and others",
    collaboration = "CYGNUS",
    title = "{CYGNO: a CYGNUs Collaboration 1 m$^3$ Module with Optical Readout for Directional Dark Matter Search}",
    eprint = "1901.04190",
    archivePrefix = "arXiv",
    primaryClass = "physics.ins-det",
    reportNumber = "INFN-19/2/LNF",
    month = "1",
    year = "2019"
}

@misc{depleteduranium,
  author       = {{World Nuclear Association}},
  title        = {Uranium and Depleted Uranium},
  year         = 2023,
  url          = {https://world-nuclear.org/information-library/nuclear-fuel-cycle/uranium-resources/uranium-and-depleted-uranium},
  note         = {Accessed: 2024-09-10}
}

@article{localGroup,
    author = "Herrera, Gonzalo and Ibarra, Alejandro and Shirai, Satoshi",
    title = "{Enhanced prospects for direct detection of inelastic dark matter from a non-galactic diffuse component}",
    eprint = "2301.00870",
    archivePrefix = "arXiv",
    primaryClass = "hep-ph",
    reportNumber = "TUM-HEP 1447/22, IPMU22-0070",
    doi = "10.1088/1475-7516/2023/04/026",
    journal = "JCAP",
    volume = "04",
    pages = "026",
    year = "2023"
}

@misc{JUNO_Fiducial_Cut,
    author = "Abusleme, Angel and others",
    collaboration = "JUNO",
    title = "{Potential to Identify the Neutrino Mass Ordering with Reactor Antineutrinos in JUNO}",
    eprint = "2405.18008",
    archivePrefix = "arXiv",
    primaryClass = "hep-ex",
    month = "5",
    year = "2024"
}

@article{SNO_Fiducial_Cut,
    author = "Lozza, V.",
    editor = "Danevich, Fedor and Tretyak, Vladimir",
    collaboration = "SNO+",
    title = "{Neutrinoless double beta decay search with SNO+}",
    doi = "10.1051/epjconf/20136501003",
    journal = "EPJ Web Conf.",
    volume = "65",
    pages = "01003",
    year = "2014"
}

@article{gamma_shielding,
  title={Gamma ray attenuation properties of common shielding materials},
  author={McAlister, Daniel R},
  journal={PG Research Foundation},
  year={2012},
url={https://www.eichrom.com/wp-content/uploads/2018/02/Gamma-Ray-Attenuation-White-Paper-by-D-M-rev-6-1-002.pdf}
}

@article{neutron_shielding,
  title={Neutron Shielding Materials},
  author={McAlister, Daniel R},
  journal={PG Research Foundation},
  year={2016},
url={https://www.eichrom.com/wp-content/uploads/2018/02/neutron-attenuation-white-paper-by-d-m-rev-2-1.pdf}
}

@article{archaeological_lead,
    author = "Pattavina, Luca and Ferreiro Iachellini, Nahuel and Tamborra, Irene",
    title = "{Neutrino observatory based on archaeological lead}",
    eprint = "2004.06936",
    archivePrefix = "arXiv",
    primaryClass = "astro-ph.HE",
    doi = "10.1103/PhysRevD.102.063001",
    journal = "Phys. Rev. D",
    volume = "102",
    number = "6",
    pages = "063001",
    year = "2020"
}

@inproceedings{HyperK,
    author = "Bian, J. and others",
    collaboration = "Hyper-Kamiokande",
    title = "{Hyper-Kamiokande Experiment: A Snowmass White Paper}",
    booktitle = "{Snowmass 2021}",
    eprint = "2203.02029",
    archivePrefix = "arXiv",
    primaryClass = "hep-ex",
    month = "3",
    year = "2022"
}

@article{DUNE,
    author = "Hewes, V. and others",
    collaboration = "DUNE",
    title = "{Deep Underground Neutrino Experiment (DUNE) Near Detector Conceptual Design Report}",
    eprint = "2103.13910",
    archivePrefix = "arXiv",
    primaryClass = "physics.ins-det",
    reportNumber = "FERMILAB-PUB-21-067-E-LBNF-PPD-SCD-T",
    doi = "10.3390/instruments5040031",
    journal = "Instruments",
    volume = "5",
    number = "4",
    pages = "31",
    year = "2021"
}

@misc{superK,
    author = "Walter, Christopher W.",
    title = "{The Super-Kamiokande Experiment}",
    eprint = "0802.1041",
    archivePrefix = "arXiv",
    primaryClass = "hep-ex",
    doi = "10.1142/9789812771971_0002",
    pages = "19--43",
    month = "2",
    year = "2008"
}

@article{AlbertaPaper,
    author = "Reynoso-Cordova, Javier and Bozorgnia, Nassim and Piro, Marie-C\'ecile",
    title = "{The Large Magellanic Cloud: expanding the low-mass parameter space of dark matter direct detection}",
    eprint = "2409.09119",
    archivePrefix = "arXiv",
    primaryClass = "hep-ph",
    doi = "10.1088/1475-7516/2024/12/037",
    journal = "JCAP",
    volume = "12",
    pages = "037",
    year = "2024"
}

@article{OutdatedSensitivity,
    author = {Schumann, Marc and Baudis, Laura and B\"utikofer, Lukas and Kish, Alexander and Selvi, Marco},
    title = "{Dark matter sensitivity of multi-ton liquid xenon detectors}",
    eprint = "1506.08309",
    archivePrefix = "arXiv",
    primaryClass = "physics.ins-det",
    doi = "10.1088/1475-7516/2015/10/016",
    journal = "JCAP",
    volume = "10",
    pages = "016",
    year = "2015"
}

@article{Kamland,
    author = "Gando, A. and others",
    collaboration = "KamLAND",
    title = "{$^7$Be Solar Neutrino Measurement with KamLAND}",
    eprint = "1405.6190",
    archivePrefix = "arXiv",
    primaryClass = "hep-ex",
    doi = "10.1103/PhysRevC.92.055808",
    journal = "Phys. Rev. C",
    volume = "92",
    number = "5",
    pages = "055808",
    year = "2015"
}

@article{KamlandDepth,
    author = "Piepke, A.",
    editor = "Law, J. and Ollerhead, R. W. and Simpson, J. J.",
    collaboration = "KamLAND",
    title = "{KamLAND: A reactor neutrino experiment testing the solar neutrino anomaly}",
    doi = "10.1016/S0920-5632(00)00928-2",
    journal = "Nucl. Phys. B Proc. Suppl.",
    volume = "91",
    pages = "99--104",
    year = "2001"
}

@misc{ChristGrant,
  author = {Christopher Grant},
  title = {Private communication},
  year = {2025}
}

\end{document}